\documentclass[aps,pre,onecolumn,nofootinbib]{revtex4-2}

\usepackage{amsmath,amsfonts,amssymb}
\usepackage{bm}
\usepackage{physics}
\usepackage{xcolor}
\usepackage{graphicx}
\usepackage{float}
\usepackage{mathtools}
\usepackage{tabularx}
\usepackage{xr-hyper} 
\usepackage{lpic,tikz}
\usepackage[utf8]{inputenc}
\usepackage[normalem]{ulem} 

\newcommand{\aer}{\alpha_{\text{er}}}
\newcommand{\ati}{\alpha_{\text{ti}}}

\definecolor{amethyst}{rgb}{0.6, 0.4, 0.85}

\newcommand{\revP}[1]{{\color{black} #1}} 

\begin{document}

\title{Rebound in epidemic control: How misaligned vaccination timing amplifies infection peaks}

\author{Piergiorgio Castioni}
\email{piergiorgio.castioni@urv.cat}

\author{Sergio G\'omez}
\email{sergio.gomez@urv.cat}

\author{Clara Granell}
\email{clara.granell@urv.cat}

\author{Alex Arenas}
\email{alexandre.arenas@urv.cat}
\email{corresponding author}

\affiliation{Departament d'Enginyeria Inform\`{a}tica i Matem\`{a}tiques, Universitat Rovira i Virgili, 43007 Tarragona, Spain}


\begin{abstract}
In this study, we explore the dynamic interplay between the timing of vaccination campaigns and the trajectory of disease spread in a population. Through comprehensive data analysis and modeling, we have uncovered a counter-intuitive phenomenon: initiating a vaccination process at an inopportune moment can paradoxically result in a more pronounced second peak of infections. This ``rebound'' phenomenon challenges the conventional understanding of vaccination impacts on epidemic dynamics. We provide a detailed examination of how improperly timed vaccination efforts can inadvertently reduce the overall immunity level in a population, considering both natural and vaccine-induced immunity. Our findings reveal that such a decrease in population-wide immunity can lead to a delayed, yet more severe, resurgence of cases. This study not only adds a critical dimension to our understanding of vaccination strategies in controlling pandemics but also underscores the necessity for strategically timed interventions to optimize public health outcomes. Furthermore, we compute which vaccination strategies are optimal for a COVID-19 tailored mathematical model, and find that there are two types of optimal strategies. The first type prioritizes vaccinating early and rapidly to reduce the number of deaths, while the second type acts later and more slowly to reduce the number of cases; both of them target primarily the elderly population.
Our results hold significant implications for the formulation of vaccination policies, particularly in the context of rapidly evolving infectious diseases.
\end{abstract}

\maketitle

\section*{Introduction}

Global pandemics of infectious diseases are one of the greatest threats that humanity faces and, as a consequence, the issue of how to control them becomes increasingly more urgent as time passes. To rise up to the challenge, traditional epidemic control offers two alternatives: non-pharmaceutical interventions (NPIs) such as school closures, quarantine, or mask mandates on the one hand, and pharmaceutical interventions such as vaccination, drugs, or treatments on the other. However, in the case of NPIs~\cite{gordon_cross-country_2021} as in that of vaccination~\cite{bootsma_effect_2007, matveeva_comparison_2023}, apparently similar implementations can lead to different outcomes, hinting at the fact that epidemic control is, at its heart, a complex problem that requires a detailed understanding of the underlying mechanisms of the epidemic.

Mathematical models have long been used to provide a framework to gather such insights~\cite{keeling_modeling_2008}. In the context of COVID-19, for instance, they have proved to be a powerful tool for the description, prediction and prevention of the ongoing pandemic~\cite{kucharski2020early, giordano_modelling_2020, kaxiras_multiple_2020, maier20, Arenas2020, wong2020modeling, jewell2020predictive, vespignani2020modelling, maier_potential_2021, bauch2021, bubar_sars-cov-2_2022}. 
Therefore, it comes as no surprise that the search for an optimal strategy for the minimization of the cases is a well-researched topic, in particular in the context of NPIs \cite{wallinga_optimizing_2010, hollingsworth_mitigation_2011, eames_influence_2014, morris_optimal_2021, balderrama_optimal_2022}. The reason behind this interest is obvious: non-pharmaceutical interventions come with high social and economic costs attached and it is therefore crucial to make them last as little as possible, possibly to contain the epidemic until pharmaceutical treatments are available. The same kind of optimization problem has been studied in relation to vaccination, for example in connection to issues of economical costs~\cite{klepac_synthesizing_2011}, vaccination rates~\cite{rodriguez-maroto_vaccination_2023}, age prioritization~\cite{bubar21}, boosters' distribution~\cite{alexander_modelling_2006, maier_potential_2021, matrajt_optimizing_2021} and NPIs-vaccination synergy~\cite{moore_vaccination_2021}. Following this literature, in this paper, we will also focus on the optimization of vaccination strategies, especially from the point of view of timing and duration. Unlike the previous works on this subject~\cite{feng_modeling_2011, lee_exploring_2017, costantino_modelling_2019, lauro_optimal_2021}, we employ an SIRS model with a time-limited vaccination campaign, which stands as the simplest model incorporating waning immunity (a feature shared by vaccines of several respiratory illnesses \cite{young_duration_2018, lipsitch_sars-cov-2_2022}). Additionally, we assess the effectiveness of a vaccination strategy by measuring the peak prevalence observed after the conclusion of the vaccination campaign.
From our analysis, a counter-intuitive rebound effect is observed: the timing of vaccination and subsequent immunity waning can synchronize with the increase in susceptible population, potentially leading to an infection peak larger than what would be expected in the absence of vaccination. Earlier attempts to model reinfections and waning immunity employing an SIRS model exist~\cite{elbasha_analyzing_2011, nill_endemic_2023}, but they primarily deal with the stability of such systems under the assumption of an unlimited supply of vaccines, and therefore no rebound was observed. 
Resurgence of infected individuals during an epidemic has been observed in the literature in particular cases: when the infectivity of the pathogen is periodic (seasonal)~\cite{feng_modeling_2011}, when vaccination is characterized by pulsating campaigns ~\cite{agur_pulse_1993}, or when the waning considers the dynamics of within-host immunity ~\cite{heffernan_implications_2009}. However, to the best of our knowledge, it has not yet been mechanistically described in a simple compartmental model depending only on the timing of the vaccination campaign.

Finally, we are interested in investigating whether the rebound effect seen in simpler models persists when examining a specific disease and a more advanced model. For this purpose, we expand upon the model by Arenas et al.~\cite{Arenas2020}, originally developed for COVID-19, to include vaccination campaigns, focusing on the effective management of vaccination timing to minimize the second peak of infections and reduce cumulative hospitalizations and deaths. The age-stratified nature of this model also allows us to introduce an additional variable alongside the timing and duration: the age priority. Using data from a recent wave of an infectious disease, our findings provide key insights for optimizing vaccination distribution, aiming to mitigate the effects of future epidemic waves and inform public health policies.

\section*{Results}

In the following, we outline our findings about what separates a good vaccination strategy from a bad one, and in particular on how to get to an optimal one. This section is divided into four parts: The first part introduces the SIRS model with vaccination and its basic features. The second part focuses on the rebound effect (i.e., the unexpected growth of cases after a vaccination campaign) and gives a mechanistic explanation on why it happens. The third part outlines the importance of timing and duration of a campaign to achieve an optimal result. Lastly, the fourth part shifts to a more complex epidemiological model and deals with the problem of identifying the Pareto-optimal solutions that simultaneously minimize both the second wave of cases/hospitalizations and overall deaths.

\subsection*{The model}
\begin{figure}[tb!]
    \centering
    \includegraphics[width = 0.425\textwidth]{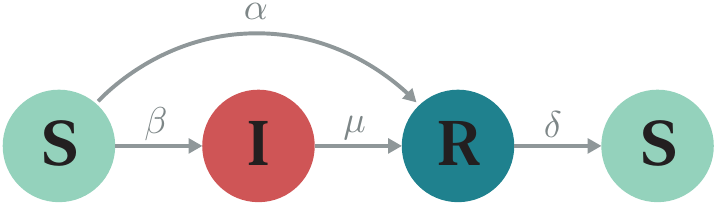}
    \caption{\textbf{SIRS model with vaccination.} A graphical representation of the SIRS compartmental model with reinfections and vaccination described in Eqs.~\eqref{eq:SIRS1}--\eqref{eq:SIRS3}.}
    \label{fig:scheme}
\end{figure}

Let us consider a continuous-time mean-field SIRS model with vaccination, a compartmental model that divides the population into three groups: susceptible (S), infected (I) and recovered (R) individuals. In this model, susceptible individuals can become infected through contact with an infected person at a rate of $\beta$. Alternatively, they may gain immunity to the disease through vaccination, occurring at a rate of $\alpha(t)$, and consequently move into the recovered category. Infected individuals transition to the recovered compartment at a rate $\mu$, while recovered individuals lose their immunity at a rate $\delta$, becoming susceptible again. Note that the vaccination process in this model can be dynamically managed; it can be turned on and off at any moment according to the employed vaccination strategy. The dynamics of the model is sketched in Fig.~\ref{fig:scheme}. \\

The equations that describe the model are: 
\begin{eqnarray}\label{eq:SIRS1}
    \dv{S}{t} &=& - \beta S I  + \delta R - \alpha(t),
    \\ \label{eq:SIRS2}
    \dv{I}{t} &=& \beta S I - \mu I,
    \\ \label{eq:SIRS3}
    \dv{R}{t} &=& \mu I - \delta R + \alpha(t).
\end{eqnarray}
Here, variables $S(t)$, $I(t)$ and $R(t)$ represent the fraction of individuals in their corresponding compartments, which fulfill $S(t)+I(t)+R(t)=1$; we also write the vaccination rate depending explicitly on time $t$, while the rest of the parameters are constant. We begin by studying the behavior of the system of Eqs.~\eqref{eq:SIRS1}--\eqref{eq:SIRS3} when $\alpha(t) = \alpha$, to understand what different levels of vaccination have on the epidemic in the long term. First of all, we recall that, if $\beta>\mu$, the stationary states $S^*, I^*, R^*$  of the standard SIRS model with no vaccination ($\alpha=0$) are \cite{wang_statistical_2016}:
\begin{equation}\label{eq:novaccine-phase}
S^* = \dfrac{1}{R_0}, \qquad I^* = \qty(1 - \dfrac{1}{R_0})\dfrac{\delta}{\mu + \delta}, \qquad R^* = \qty(1 - \dfrac{1}{R_0})\dfrac{\mu}{\mu + \delta}.
\end{equation}
The basic reproduction number of this model is $R_0=\frac{\beta}{\mu}$ and does not change once we add a non-zero vaccination term. In the case $\beta<\mu$ the system rapidly reaches the ``disease-free equilibrium'', where $S=1$ and $I=R=0$, which we consider uninteresting for the scope of this paper. Therefore, from now on we will always assume that $\beta>\mu$.

\begin{figure}[tb!]
    \centering
    \includegraphics[width = 0.65\textwidth]{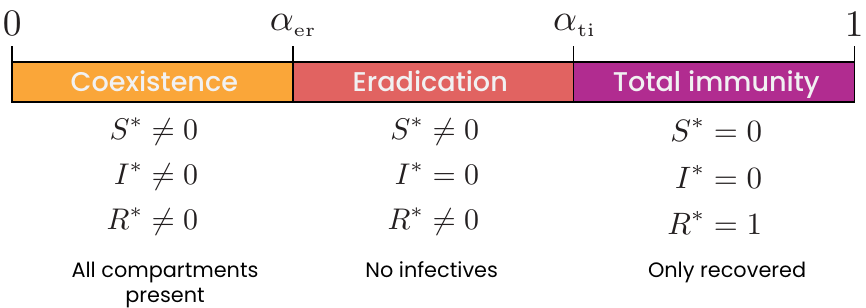}
    \caption{\textbf{Vaccination regimes.} Illustration of the different vaccination regimes based on the value of the vaccination rate~$\alpha$. Please note that the three regimes are shown with identical lengths in this diagram, though their actual extents are governed by the values of~$\aer$ and~$\ati$.}
    \label{fig:regimes}
\end{figure}

Depending on the rate of vaccination $\alpha$, we can identify three phases: the \textit{coexistence} phase, the \textit{eradication} phase, and the \textit{total immunity} phase, illustrated in Fig.~\ref{fig:regimes}. In the first phase, which occurs when the vaccination rate $\alpha\in \qty[0, \aer]$, all compartments are present and the qualitative behavior of the system remains the same as the standard SIR but with stationary values $S^*_{\text{co}}$, $I^*_{\text{co}}$, $R^*_{\text{co}}$:
\begin{equation}\label{eq:coexistence-phase}
S^*_{\text{co}} = \dfrac{1}{R_0}, \qquad I^*_{\text{co}} = I^* - \dfrac{\alpha}{\mu + \delta},
\qquad R^*_{\text{co}} = R^* + \dfrac{\alpha}{\mu + \delta}.
\end{equation}
The critical vaccination rate for eradication $\aer = \qty(1-1/R_0)\delta$ is defined as the value of the vaccination rate for which the endemic fraction of infected individuals $I^*$ goes to zero. Thus, the eradication phase, a phase where no infective individuals are present, happens when $\alpha \in \qty[\aer, \ati]$, where the stationary values become
\begin{equation}\label{eq:eradication-phase}
S^*_{\text{er}} = 1 - \dfrac{\alpha}{\delta}, \qquad I^*_{\text{er}} = 0, \qquad R^*_{\text{er}} = \dfrac{\alpha}{\delta}.
\end{equation}
The critical vaccination rate required to achieve total immunity, denoted $\alpha_{\text{ti}} = \delta$, is defined as the vaccination rate at which the fraction of susceptible individuals also vanishes. If $\alpha$ exceeds this threshold, the system transitions into the third phase—the total immunity phase—characterized by the presence of only recovered individuals. \revP{Finally, during all of our numerical simulations we never allow the fraction of susceptibles to drop lower than 0.005 due to vaccination, in order to avoid pathological behaviors with unrealistically small numbers for the fraction of individuals in any compartment.}

\subsection*{The rebound effect of misaligned vaccination}

In the previous section, we assumed that the rate of vaccination of the SIRS model is constant over time; however, this is far from a realistic scenario. In many real-world emergency situations, vaccine supplies are constrained and distributed within specific timeframes. Consequently, vaccination efforts extend over defined periods known as \textit{vaccination campaigns}. These campaigns have a designated start date and follow a daily schedule for distributing vaccine doses.

Building upon this premise, to incorporate a vaccination campaign into the previous model, we propose to use a piecewise function for the vaccination rate $\alpha(t)$. This function is defined for the duration of the campaign, with a starting time denoted as $t_\text{start}$ and an ending time denoted as $t_\text{stop}$, as follows:

\begin{equation}
    \alpha(t) = \begin{cases}
        \alpha \quad \text{if }t\in\qty[t_\text{start}, t_\text{stop} ],
        \\
        0 \quad \text{otherwise}.
    \end{cases}
\end{equation}

\begin{figure}[tb!]
    \centering
    \includegraphics[width = 0.99\textwidth]{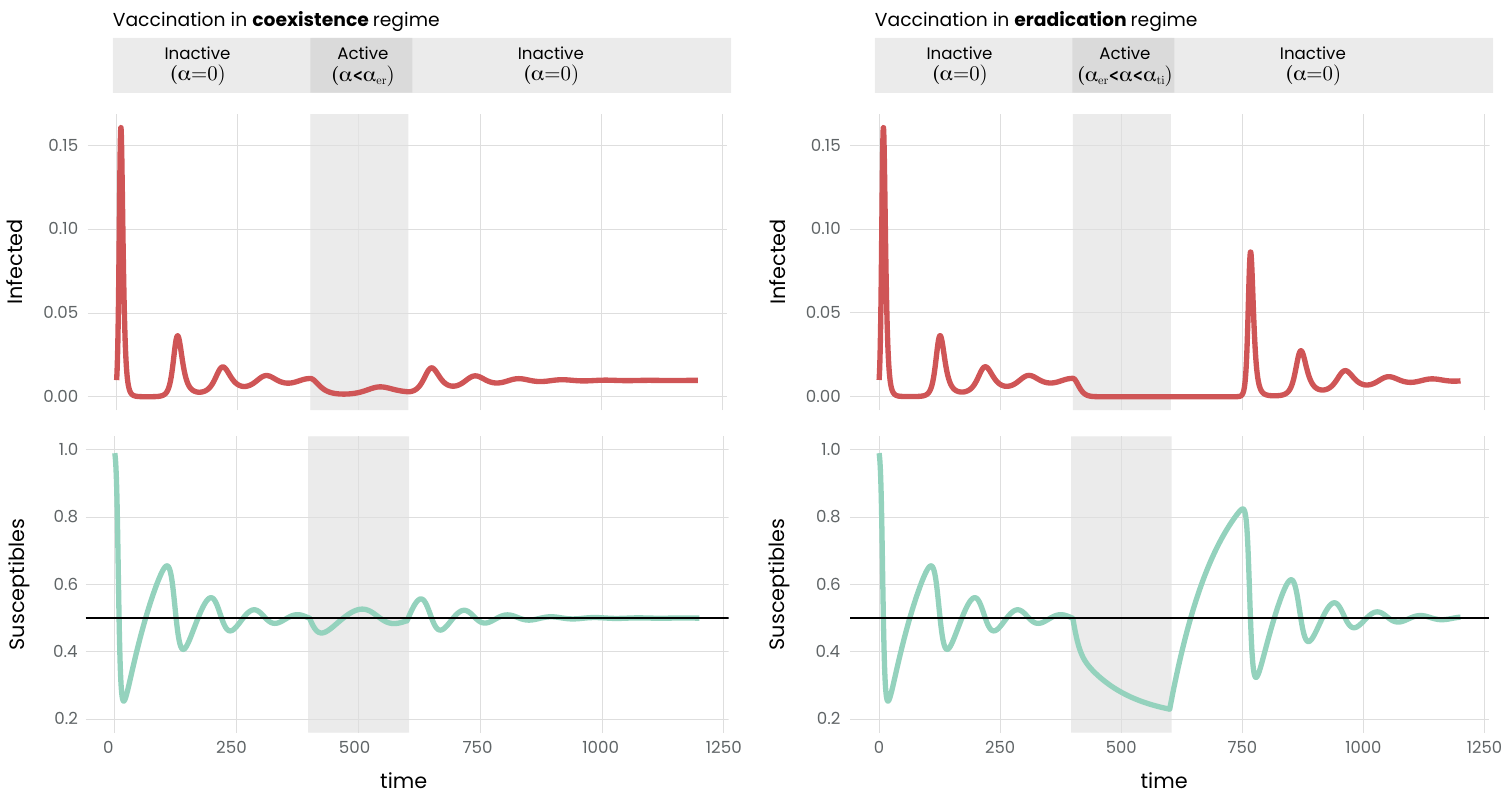}
    \caption{\textbf{Rebound of infections after vaccination in the coexistence and eradication regimes.} Illustrative example of the effects of limited-time vaccination in two different vaccination regimes. On the left, the vaccination rate $\alpha$ falls in the coexistence regime ($\alpha = 0.003$), while on the right it is in the eradication regime ($\alpha = 0.008$). The shaded area denotes the time period when the vaccination strategy is active. In both regimes, when the vaccination stops, a resurgence wave is observed, but with a different magnitude: in the coexistence regime, the wave is relatively modest, whereas in the second scenario, it is more pronounced, comparable in size to the initial wave. This observation suggests that a higher intensity of vaccination efforts may paradoxically exacerbate the rebound effect. Simulation with $\beta=1$, $\mu=0.5$, $\delta = 0.01$ and a vaccination campaign taking place between $t=400$ and $t=600$. The initial conditions are $I(0)=0.01$ and $S(0)=0.99$. }
    \label{fig:example-rebound}
\end{figure}

Let us begin our analysis by simulating the previous epidemic model using two different values for the vaccination rate $\alpha$, thereby placing the model in two distinct vaccination regimes\footnote{Throughout all our numerical simulations, we maintain the fraction of susceptibles above 0.005 following vaccination to prevent pathological behaviors associated with unrealistically small numbers in any compartment.}. Concerning timing, we will let the model reach a stable equilibrium before initiating the corresponding vaccination campaign. We simulate the system of Eqs.~\eqref{eq:SIRS1}--\eqref{eq:SIRS3} with a value of $\alpha=0.003$ (corresponding to the coexistence phase) and $\alpha=0.008$ (corresponding to the eradication phase), see Fig.~\ref{fig:example-rebound}.

In the figure, we observe that initially, without any vaccination strategy in place, the system experiences fluctuations until reaching a stable equilibrium. Following this, the initiation of the vaccination campaign leads to a reduction in the number of infected individuals. This reduction also involves the susceptibles if the strategy falls under the eradication regime.

Then, after ending the vaccination campaigns, we observe the anticipated ``\textit{rebound effect}'', characterized by a significant peak in the number of infected individuals, which is particularly evident in the eradication regime. This phenomenon hints that vaccination, despite its benefits, can have complex and sometimes unexpected effects on disease spread dynamics, and deserves a deeper investigation into its origins and underlying mechanisms.

This counter-intuitive phenomenon, like many aspects of epidemic modeling, can be understood through the availability of susceptible individuals. A disease can only spread if there are susceptible individuals to fuel its transmission. In this light, the efficacy of a vaccination campaign lies in its ability to transfer people directly from the susceptible category to the recovered one, thereby reducing the pool of individuals available for contagion. However, once the campaign concludes and assuming vaccine-induced immunity wanes over time, vaccinated individuals gradually revert to susceptibility. If the campaign was highly effective, a significant portion of the population might become susceptible simultaneously, lacking natural immunity from recent infections and becoming available for new infections. This synchrony can lead to an increase in infection numbers. Analogous to pulling and releasing a spring, the vaccination campaign initially lowers infection rates, but their subsequent increase can lead to a sharp rise in cases.

It is important to note that the extent of this rebound varies significantly depending on the vaccination strategy employed. In the co-existence regime, the rebound is small and can simply be seen as the system settling from one equilibrium (with vaccines, Eq.~\eqref{eq:coexistence-phase}) to another (without vaccines, Eq.~\eqref{eq:novaccine-phase}). On the other hand, in the eradication and total-immunity regimes, the rebound can be very large (depending on the duration of the vaccination campaign). 

To mathematically understand the mechanism behind the rebound in the latter case, we must return to the modeling approach. As we mentioned, the eradication regime starts at the value of $\alpha_\text{er}$ where the number of infected individuals goes to zero. This allows us to develop a perturbation theory framework that gives us an approximated solution for the unfolding of the epidemic just after the end of the vaccination campaigns. At zeroth order in this approximation the contagion term $\beta S I$ and the recovery term $\mu I$ vanish so the system of Eqs.~\eqref{eq:SIRS1}--\eqref{eq:SIRS3} can be approximated by:
\begin{eqnarray}\label{eq:SR1}
    \dv{S_0}{t} &=&   \delta R_0,
    \\ \label{eq:SR2}
    \dv{R_0}{t} &=&   - \delta R_0,
\end{eqnarray}
whose analytical solutions are governed by exponential functions. The full calculations at first order together with all the definitions can be found in the Methods section. The full solution for $t \geqslant t_\text{stop}$  then reads:
\begin{eqnarray}
  \label{eq:analytica_sol1}
  S(t) &=& 1 - R(t_\text{stop})e^{-\delta (t- t_\text{stop})} + \widetilde{\Delta S}(I(t), t),
\\
  \label{eq:analytica_sol2}
  R(t) &=& R(t_\text{stop})e^{-\delta (t - t_\text{stop}) } + \widetilde{\Delta R}(I(t), t),
\\
  \label{eq:analytica_sol3}
  I(t) &=& I(t_\text{stop}) \exp\left[ (\beta -\mu) (t-t_\text{stop}) - \frac{\beta R(t_\text{stop})}{\delta}(1- e^{-\delta (t-t_\text{stop})})\right],
 \end{eqnarray}
where $\widetilde{\Delta S}(I(t), t)$, $\widetilde{\Delta R}(I(t), t)$ and $I(t)$ are the first-order contributions, and $S(t_\text{stop})$, $R(t_\text{stop})$ and $I(t_\text{stop})$ are the value for each compartment right after the end of the vaccination process  (the agreement between these equations and the numerical solution is shown in Fig.~S1 of the Supplementary Material). It is important to note that these equations are applicable only for a certain period after $t_\text{stop}$. Indeed, as long as the number of infected $I(t)$ stays low, the contributions of $\widetilde{\Delta S}(I(t), t)$ and $\widetilde{\Delta R}(I(t), t)$ also remain small, allowing for the zeroth order of Eqs.~\eqref{eq:analytica_sol1}--\eqref{eq:analytica_sol2} to dominate and the susceptible compartment to approach its limit $S\rightarrow 1$.

Despite their limitations, these equations provide valuable insights and allow us to draw a causal connection between the strength of the vaccination campaign and the height of the rebound. For instance, we know that the approximation stops working at the time $t_p$ where the number of infected, Eq.~\eqref{eq:analytica_sol3}, reaches its theoretical upper bound, i.e., $I(t_p) = 1 - \mu / \beta$. A closed, although complicated, expression for $t_p$ exists and thanks to that it is still possible to demonstrate that $\pdv{t_p}{S(t_\text{stop})}<0$ and $\pdv{t_p}{I(t_\text{stop})}<0$ (see proof in Section~S1 of the Supplementary Material). This implies that the smaller the number of susceptible and infective individuals remaining after vaccination, the larger $t_p$ will be; thus, the validity of the exponential solutions to Eqs.~\eqref{eq:analytica_sol1}--\eqref{eq:analytica_sol2} extends over a longer time period. Since $t_p$ is connected to the growing phase of the zeroth order susceptible in Eq.~\eqref{eq:analytica_sol1}, a larger $t_p$ also means a larger build-up in the susceptible compartment and therefore a larger peak of susceptibles $S_\text{max}$. Finally, one can prove, albeit based on other simplifying assumptions (see Section~S2 of the Supplementary Material), that a larger $S_\text{max}$ results in a taller epidemic peak (i.e., the height of the first peak of infectives following vaccination). This, therefore, demonstrates that a robust vaccination campaign can cause the rebound peak to be taller. 

\subsection*{Timing as a fundamental feature to avoid the rebound}

In the preceding section, we exclusively examined scenarios where the vaccination campaign commenced after the standard SIRS model had reached its equilibrium. Furthermore, the campaign had a designated duration intended to allow the system enough time to stabilize into a new equilibrium state under the influence of vaccination before returning to its initial condition upon the campaign's completion.
However, vaccination campaigns  frequently overlap with epidemic waves that are still far from equilibrium, and their durations may differ. This highlights the importance of identifying the optimal timing and duration of vaccination campaigns to reduce the rebound effect while the epidemic is still ongoing.

To address this challenge, we have developed a simulation framework to systematically evaluate a broad spectrum of vaccination strategies, spanning various starting times and durations. 
Our objective is to find the strategy that most effectively counters the rebound effect by specifically aiming to minimize the height of the next largest peak\footnote{The notion of next largest peak usually coincides with that of the second peak, but that is not always the case, as shown in Fig.~S4 in the Supplementary Material.}, which defines our optimal criterion in this context. 

\begin{figure}[tb!]
    \centering
    \includegraphics[width = 0.8\textwidth]{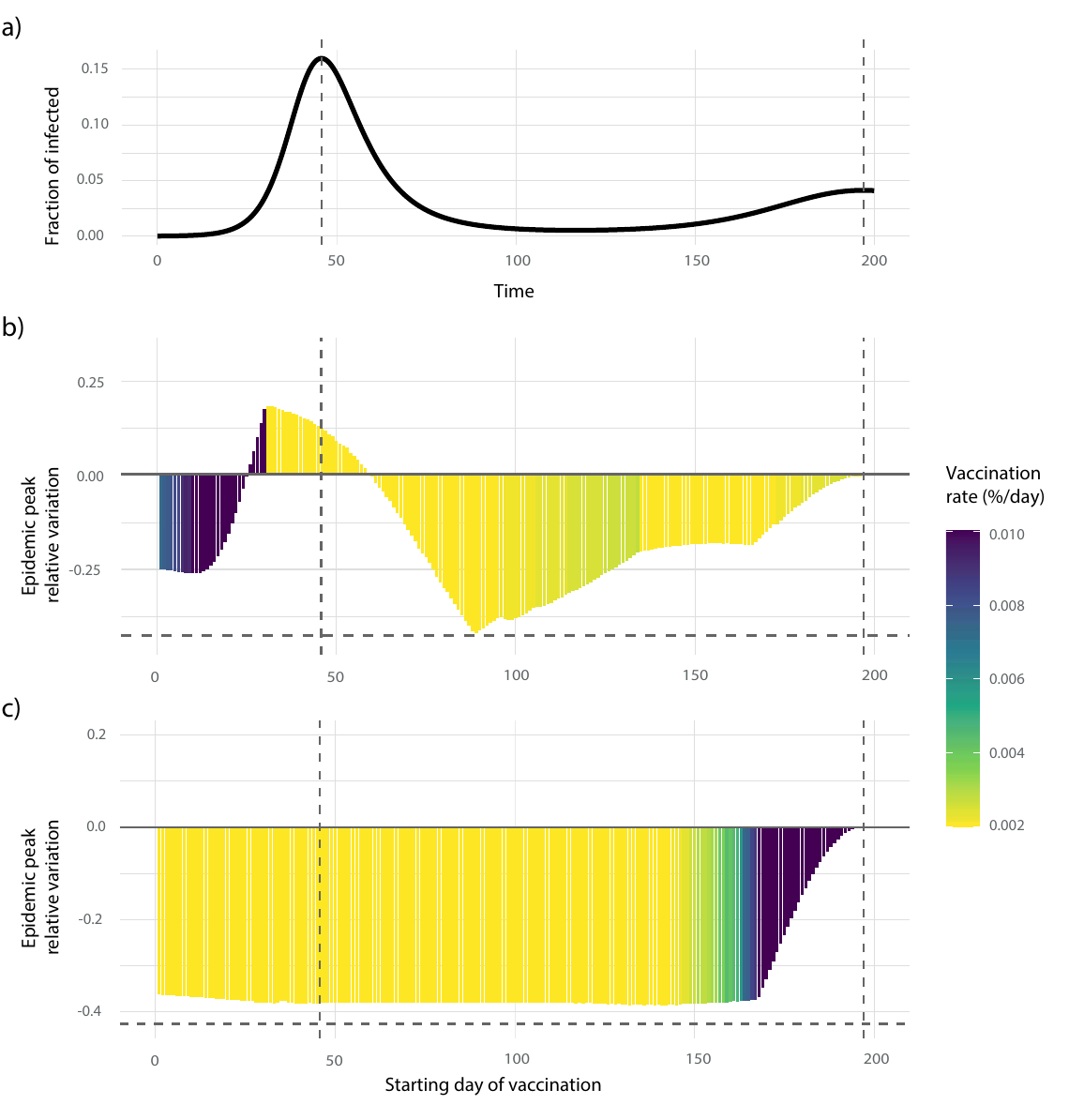}
    \caption{\textbf{Best vaccination strategy for each possible starting day before the second peak.} Panel a) illustrates the proportion of infectious individuals over time in the baseline scenario, i.e., without vaccination. The vertical dashed lines indicate the peaks of the epidemic and act as reference points in the subsequent panels.
    Panel b) shows the variation in the height of the subsequent largest peak relative to the baseline (vertical axis) due to the most effective vaccination strategy initiated on each day (horizontal axis). Here, the term ``best'' strategy refers to the approach that results in the greatest reduction in the height of the subsequent largest peak. The color of each bar indicates the vaccination rate of that strategy.
    Panel c) shows the same quantities as panel b) but instead of using a constant vaccination rate $\alpha$ we have used vaccination rate $\alpha \propto \Theta(S-S^*)$. 
    The horizontal black dashed lines in panels b) and c) correspond to the maximum possible reduction, obtained when the second peak is just as tall as the equilibrium value. In this case, the strategy that comes closer to achieve the optimal value is the one in panel b) that starts on day $89$ with speed $\alpha=0.002$. }
    \label{fig:scenarios_SIRS}
\end{figure}

To achieve this, we initially conduct a simulation without any vaccination to establish a baseline that serves as a reference point for comparing all subsequent vaccination scenarios. In simulations that incorporate vaccination, we preserve the baseline scenario's parameters and initial conditions. A vaccination campaign is characterized by a specific \textit{starting day} and \textit{duration}. Note that the duration of the campaign is inversely proportional to the rate of vaccination $\alpha$. We perform a separate simulation for each potential start day, ranging from day~1 to the day when the baseline scenario reaches its second peak (indicated by the second vertical line in Fig.~\ref{fig:scenarios_SIRS}(a)).

Additionally, we vary the duration of the vaccination campaign, exploring periods between~30 to 150~days. In all simulations, the total number of vaccines is fixed, equal to~30\% of the population. This quantity ensures that, under the fastest rollout scenario of 30~days, the daily vaccination rate reaches~1\% of the population. We select this rate as the maximum feasible daily coverage in  a realistic scenario. The decision to use a fixed total quantity of vaccines was driven by our aim to optimize the use of a resource we consider to be finite.

Our goal is to identify the strategy that maximizes the benefits derived from the limited number of vaccines available. This approach is based on the understanding that having access to additional vaccines would be advantageous, as a greater quantity of vaccines generally leads to more favorable outcomes. This assumption is corroborated by the data presented in Fig.~S2 in the Supplementary Material, which illustrates the inverse relationship between the total number of vaccines and the cumulative number of cases. However, it is critical to recognize that merely increasing the vaccine supply does not unequivocally resolve all challenges. Specifically, as the vaccine allocation escalates, the rebound effect may become more pronounced, as evidenced in Fig.~S3 of the Supplementary Material. This observation highlights the need to develop strategies that not only make the most of the existing vaccine supply but also minimize the risk of a rebound effect. By doing so, we can improve the overall effectiveness of vaccination campaigns in controlling epidemic outbreaks.

Our findings are illustrated in Fig.~\ref{fig:scenarios_SIRS}(b), where we present the outcomes of  vaccination campaigns that start on different days. Each bar in these plots corresponds to a specific start day for the vaccination, with the bar's color indicating the campaign's duration and the height indicating the relative reduction or increase in the size of the subsequent peak compared to the baseline scenario. A bar extending upwards signifies an increase in the next peak, indicating a rebound effect, whereas a downward extension suggests a reduction in peak size, indicating no rebound effect. It is important to note that although multiple scenarios with varying durations were analyzed for each start day, the plots only display the outcome of the optimal strategy, defined as the one resulting in the smallest subsequent peak.

From our analysis, we find that if the vaccination roll-out starts around the time of the first peak, even the most successful vaccination strategies will suffer from the rebound effect. This can be seen in Fig.~\ref{fig:scenarios_SIRS}(b), where we see that those simulations that consider a starting time around the first peak experience the rebound effect, indicated by bars with positive heights, signifying an increase in the epidemic's peak magnitude. In turn, strategies that start after the first peak have a better chance of avoiding any resurgence, particularly when vaccines are administered over an extended period (i.e., with vaccination within the coexistence regime). The optimal strategy observed occurs just a few days following the first peak and is the only approach that achieves a complete flattening of the infectious curve. This is evident from its proximity to the dashed line in Fig.~\ref{fig:scenarios_SIRS}(b), which indicates a reduction making the second peak equivalent to the equilibrium value. Outside of this optimal strategy, determining a straightforward rule for how soon before the second peak to start and the ideal length of the campaign proves challenging.

To test the idea introduced in the previous section, i.e., that the rebound effect is caused by an excessive departure of the susceptible and infected from their equilibrium values $S^*$ and $I^*$, we introduce a new kind of vaccination campaign. Previously, our model used constant, nonzero vaccination rates $\alpha(t)$ during a specific time frame, with rates dropping to zero outside of this interval. To avoid pulling the number of susceptible too far away from their equilibrium and thus cause the rebound effect, we test a new vaccination strategy. Here we introduce a new formula for the vaccination rate $\alpha(t, S) \propto \Theta(S-S^*)$, where $\Theta$ indicates the Heaviside function. This functional form ensures that the vaccination will only be active when the susceptibles are above their equilibrium number. The vaccination finally stops at the moment in time when all the doses at our disposal (30\% of the population in this case) run out. Please note that this strategy is characterized by intermittent activation, as the susceptible population's numbers oscillate around the equilibrium point. This fluctuation causes the vaccination campaign to switch on and off, until it finally ceases when the supply of vaccines is fully depleted.
It can be seen in Fig.~\ref{fig:scenarios_SIRS}(c) that this ``fine-tuned'' strategy works wonderfully, resulting in a consistent relative reduction across a wide range of starting times and almost always outperforming the constant strategy. The only exception to this rule are the truly optimal solutions leading to the total flattening of the second peak, that can only be obtained with the constant strategy.

\subsection*{Adding complexity: rebound effect in an age-stratified metapopulation model}

In the preceding section, we explored how the rebound effect is influenced by two factors: the vaccination start day and its duration. To verify the rebound effect's consistency in more complex scenarios, we transition to the model by Arenas et al.~\cite{Arenas2020}, which was developed to simulate the spread of COVID-19 during its initial outbreak. A brief overview of this model's key components is provided here, with a comprehensive explanation available in the Methods section and further details in Sections~S3 and~S4 of the Supplementary Material.

\begin{figure}[tb!]
    \centering
    \includegraphics[width = 0.8\textwidth]{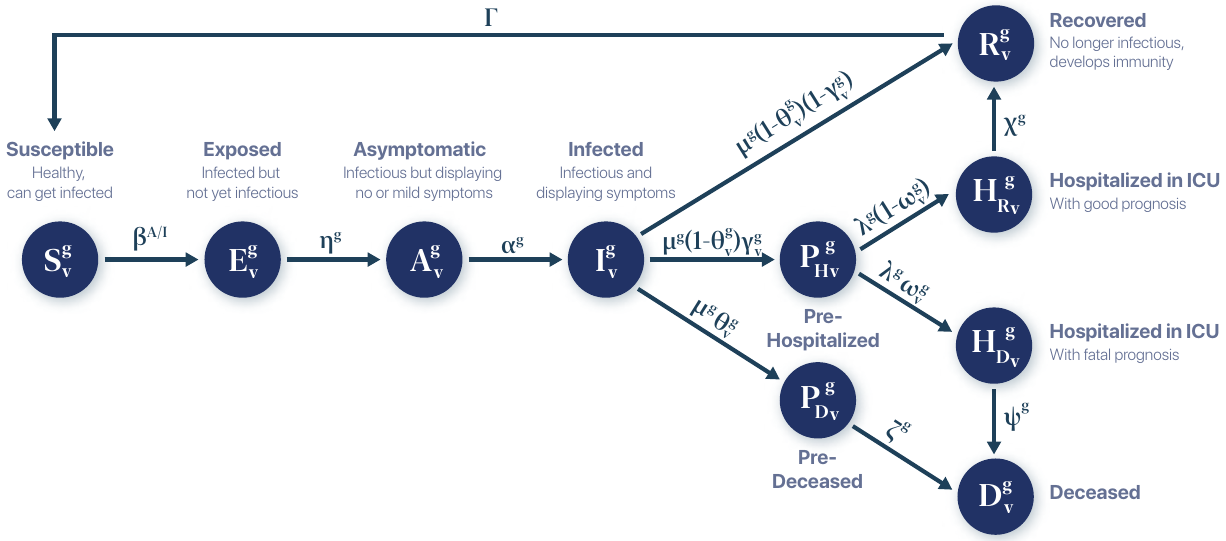}
    \caption{ \textbf{Compartments of the epidemic model for COVID-19.} The acronyms correspond to susceptible ($S^g_v$), exposed ($E^g_v$), asymptomatic infectious ($A^g_v$), symptomatic infectious ($I^g_v$), pre-hospitalized in ICU ($P^g_{Hv}$), pre-deceased ($P^g_{Dv}$), in ICU before recovery ($H^g_{Rv}$), in ICU before death ($H^g_{Dv}$), deceased ($D^g_v$), and recovered ($R^g_v$), where $g$ denotes the age stratum of all compartments and $v$ the vaccination status. The arrows indicate the transition probabilities.}
    \label{fig:MMMCA_sketch}
\end{figure}

Building upon the conventional SIRS framework, this model integrates further phases of the infection cycle, notably the exposed (E) and asymptomatic (A) stages, in addition to hospitalization (H) and death (D). Additional compartments were introduced to accurately represent the timeframes leading to hospitalization and death. An illustration of the compartmental model is provided in Fig.~\ref{fig:MMMCA_sketch}. The model employs a metapopulation approach, incorporating the movement-interaction-return mechanism~\cite{gomez2018critical}. This implies that the population is distributed across various patches, with individuals moving to adjacent areas and then returning to their original locations, without permanent migration. Moreover, the model incorporates age stratification, categorizing individuals as young, adult, or old, to account for the infection's outcome varying significantly with age. However, the original model in~\cite{Arenas2020} does not account for vaccination and waning immunity. Therefore, we adapted the model to include these aspects, as shown in Fig.~\ref{fig:MMMCA_sketch}.

As with the SIRS model, we start by running a baseline simulation without vaccination (see Fig.~S5 in the Supplementary Material). Then, we run an extensive number of simulations where we systematically explored various combinations of durations, starting days, and age priorities, all while maintaining a constant total number of vaccines. We vary the starting day of the vaccination, considering all days between the beginning of the simulation and the midpoint between the first and second peak. For the duration of the vaccination, we use 30, 60, 90, and 120~days and we keep the total number of doses fixed (so that all strategies deliver the same amount of vaccines). To explore the age group priority, we used all the possible combinations of strictly positive values that are multiples of $10\%$ and that sum up to $100\%$. The resulting strategies are then categorized as young, adult, old, or mixed based on the priority levels of each age group. For instance, a strategy labeled ``young'' means that the priority given to the young group is higher than any of the priorities given to the two remaining groups. The mixed strategies are those that do not fulfill the previous condition. 

\begin{figure}[tb!]
    \centering
    \includegraphics[width = 0.9\textwidth]{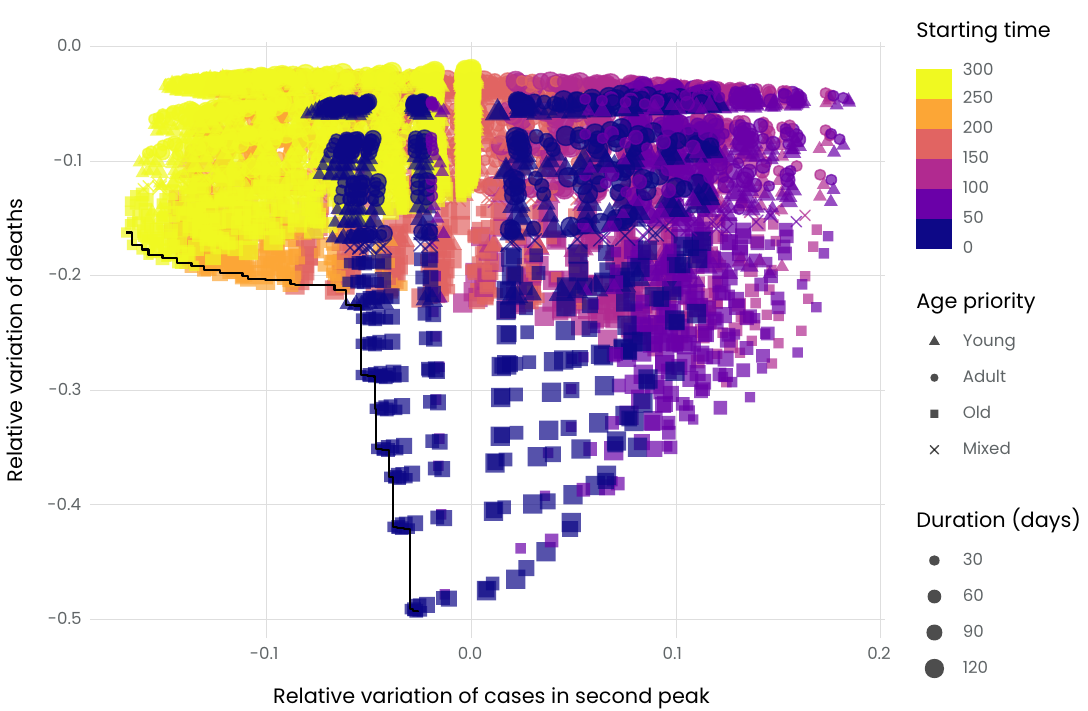}
    \caption{\textbf{Analysis of the vaccination strategies.} In this scatter plot, each point corresponds to a unique simulation based on a distinct vaccination strategy. The horizontal axis displays the relative variation in the number of cases at the next largest peak relative to the baseline case, while the vertical axis represents the relative variation in the number of cumulative deaths, both with respect to the baseline. Our objective is to minimize these two metrics in order to identify strategies that outperform the baseline scenario. Each plot point conveys three distinct variables: color, shape, and size. The color signifies the strategy's implementation time ($t_\mathrm{start}$), with brighter shades indicating later starting time. The point's shape indicates the primary age group targeted by the strategy, and the point's size reflects the campaign's duration, where larger sizes denote longer duration. It is important to note that the total number of vaccines administered remains constant, irrespective of the campaign's duration. As we can see, many of the analyzed strategies yield negative outcomes. Points on the positive side of the horizontal axis indicate an increase in the number of cases during the second peak. However, it is important to note that none of the simulations results in an increase in the number of deaths. Finally, the black solid line marks the Pareto front, which indicates the optimal strategies.}
    \label{fig:pareto}
\end{figure}

\begin{figure}[tb!]
    \centering
    \includegraphics[width = 0.95\textwidth]{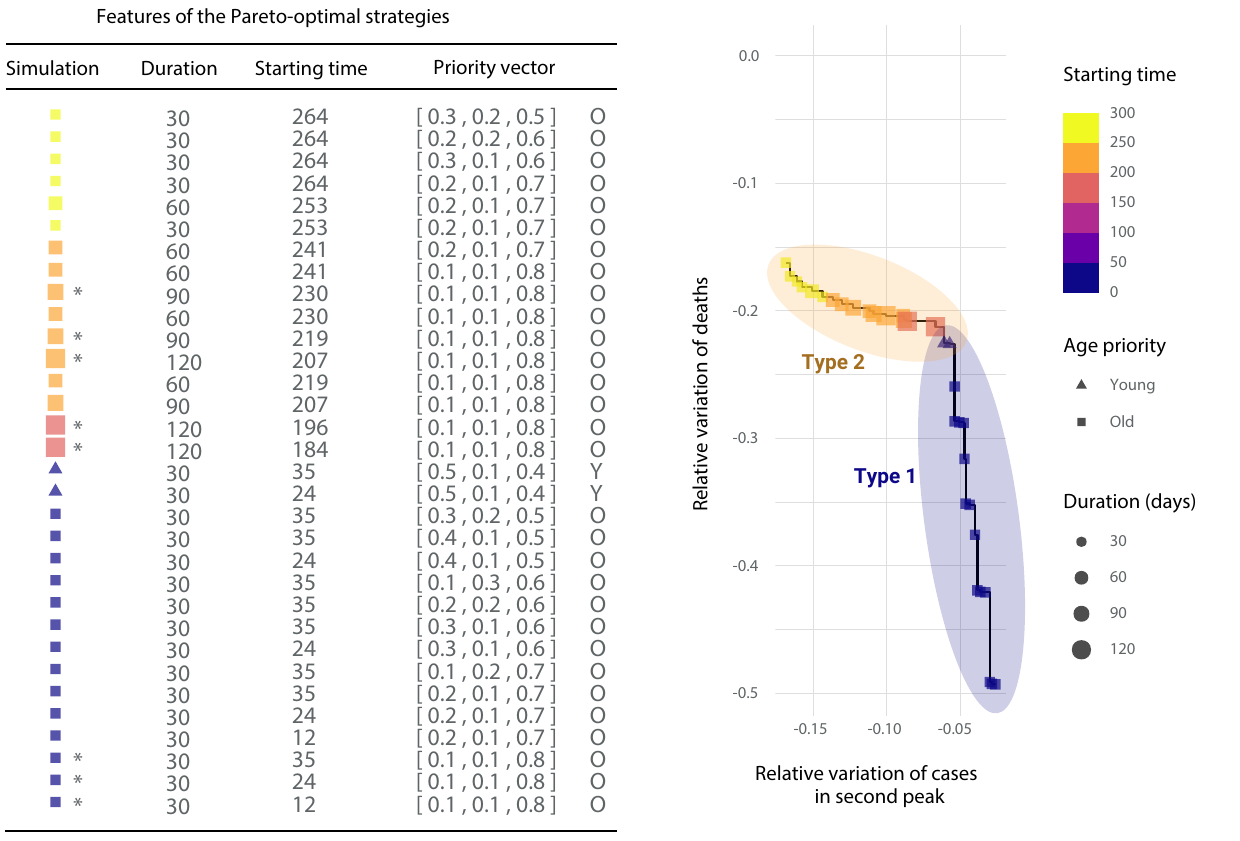}
    \caption{\textbf{Features of the Pareto-optimal strategies.} This table shows the characteristics of Pareto-optimal strategies alongside a scatter plot that includes only those data points belonging to the Pareto front. The columns of the table provide information on the duration and starting time of each strategy, the priority vector that determines the vaccines distribution by age group (ordered as [Young, Adult, Old]), as well as the age stratum that is most prioritized for each strategy. The asterisk indicates simulations that are optimal for reducing both the number of cases and the number of hospitalizations (see Fig.~S7). It is possible to identify two different sets of strategies, highlighted by yellow and violet ovals and labeled Type~1 and Type~2. In Type~1 strategies, the primary objective is to minimize the number of deaths, indicated by the lower values on the vertical axis. This is best achieved by vaccinating early, before the first peak, and using short-duration campaigns. Conversely, Type~2 strategies aim to reduce the number of cases, indicated by the lower values on the horizontal axis. To achieve this goal, it is optimal to vaccinate later, before the second peak, and use longer duration campaigns. Both types suggest to strongly prioritize elderly people over everyone else.}
    \label{fig:pareto-front-table}
\end{figure}
This time, to determine the optimal strategy, we employ two metrics: the relative variation of the height of the second peak, as in the previous section, and the relative variation of cumulative deaths compared to the baseline simulation. Our objective is to minimize both metrics simultaneously by identifying Pareto-optimal solutions. These are solutions where improvements in one metric would lead to worsening the other. The latter metric, introduced due to the inclusion of age structure in our model, accounts for varying risk levels across different age groups, making it crucial for evaluating the impact of prioritizing one age group over another in vaccination plans. However, this metric is significantly affected by the total runtime of the simulation. To circumvent arbitrary decision-making, we conclude the simulation after the epidemic's transient phase.

The results are shown in Fig.~\ref{fig:pareto}. It can be seen that most of the simulations (each represented by a point) fall on positive values of the difference in cases, meaning that they perform poorly or are counterproductive, so we restrict our attention to those solutions that are able to improve the baseline scenario. Among those, the optimal ones are the ones that satisfy the aforementioned Pareto condition. Such points lay at the left-most side of the scatter plot and taken together lay on a line called the Pareto front, which we highlight in Fig.~\ref{fig:pareto-front-table}, where we also show a detailed description of the features of each of these optimal solutions. From this figure, it emerges that there are two kinds of optimal strategies, each distinctly different from the other. The first type, that we mark as Type~1, contains all those strategies that are better at reducing the number of deaths while overlooking the height of the next largest peak. In order to achieve this the vaccination has to start before the first peak is reached, it must be fast and must prioritize old people. On the other hand, the second type performs better at lowering the height of the next largest peak while allowing for a larger number of deaths. In this second case the vaccination has to start well after the first peak, must have a longer duration, and should also prioritize old people.

An analogous analysis optimizing the number of hospitalizations instead of cases can be found in Figs.~S6 and~S7. This analysis demonstrates similar behavior, as the strategies are divided into the same two groups. However, the transition between Type~1 and Type~2 strategies is much smoother and does not present the characteristic right angle that we see in Fig. \ref{fig:pareto}. This can be attributed to the fact that hospitalizations' peak and cumulative deaths are closely connected and therefore the strategies to optimally reduce the first are bound to have a deep impact on the second as well.

\section*{Discussion}

From our analysis, it emerges that the outcome of a vaccination campaign, assuming vaccine waning and immunity decay, is more strongly influenced by the timing of its initiation than by any other factor. This finding is in agreement with previous studies \cite{feng_modeling_2011, lauro_optimal_2021}. Even when we have a large number of vaccines at our disposal, a poor choice in the vaccination timing could lead to counterproductive effects. This happens because two forces are at play: on the one hand, vaccines have a high impact in the short term, helping to reduce the number of cases and fatalities; on the other hand, if the vaccination reduces the infections well below their equilibrium value, that can provoke a rise in the susceptible population and a subsequent increase in the severity of subsequent waves. This rebound effect is the main obstacle to an efficient vaccination strategy and it is therefore a phenomenon worth studying. For this reason in this work we investigated this effect by giving it a mechanistic explanation based on an approximation of the SIRS model. In this regard, all the simulations hint to the fact that the best time to start vaccinating is before the cases start rising, i.e., either before or after the epidemic peak. In particular, if the objective is to avoid the rebound completely, a fined-tuned strategy, specifically created to keep the number of susceptible people close to their equilibrium value, is the preferable choice. Finally, in a more complex scenario, different strategies are available depending on the aim of the campaign and of time at our disposal: to minimize deaths, the priority should be given to fast and early strategies; if instead our goal is to lower the peak prevalence, slower and later strategies are the better solution.

\section*{Methods}

\subsection*{No-infected approximation of the SIRS model}
As already mentioned, once a vaccination campaign in the eradication regime comes to an end, it is possible to find an approximate solution for the evolution of the system. That happens because in that timeframe the number of infectious individuals is so small that the system starts evolving as if there were no infections nor recoveries and the whole dynamics was dominatated by the waning immunity. That means that the system is evolving according to the no-infected equations in \eqref{eq:SR1} and \eqref{eq:SR2}, whose solution are
\begin{eqnarray} \label{eq:no-infected1}
    S_0(t) &=& 1 - R(t_\text{stop}) e^{-\delta (t-t_\text{stop})} \\ \label{eq:no-infected2}
    R_0(t) &=& R(t_\text{stop}) e^{-\delta (t-t_\text{stop})}
\end{eqnarray}
where $S_0(t) + R_0(t) = 1$, since we explicitly assume that $I_0(t) = 0$. The subscript $0$ has been introduced to differentiate the solution of this simplified system from those of the full SIRS model. Furthermore, we can introduce the quantities $\Delta S(t) \equiv S(t) - S_0(t)$ and $\Delta R(t) \equiv R(t) - R_0(t)$, which are the differences between the solutions of the full SIRS model and the solutions of the infected-less system. Both of these terms are of order $O(I)$, therefore as long as $I(t)$ stays small they will also remain small. Finally it should be noted that since we know that $S(t)+I(t)+R(t)=1$ and $S_0(t) + R_0(t) = 1$ it follows that $\Delta S(t) + I(t) + \Delta R(t) = 0$.

By rewriting Eq.~\eqref{eq:SIRS2} using the definition of $\Delta S(t)$ and dropping the term which is of order $O(I^2)$, we can obtain an approximated equation for the infected compartment $I(t)$
\begin{eqnarray}
    \dv{I}{t} &=& \beta S(t) I(t) - \mu I(t) \\
    &=& \beta (S_0(t) + \Delta S(t)) I(t) - \mu I(t) \\
    &=& \beta S_0(t) I(t) - \mu I(t) + O(I^2)
\end{eqnarray}
The solution to this equation is found through the method of separation of variables and is given in Eq.~\eqref{eq:analytica_sol3}. By repeating the same process with Eq.~\eqref{eq:SIRS3} we can then use the newly found $I(t)$ to find a first order estimation of the function $\Delta R(t)$, which we call $\widetilde{\Delta R}(I(t), t)$
\begin{eqnarray}
    \dv{\widetilde{\Delta R}}{t} &=& \mu I(t) - \delta \widetilde{\Delta R}(t)
\end{eqnarray}
This equation can be solved by recognizing that it is a linear equation where $I(t)$ plays the role of a forcing term. The solution is:
\begin{equation}
    \widetilde{\Delta R}(I(t),t) = \Delta R(t_\text{stop}) e^{-\delta (t-t_\text{stop})} + \mu \int_{t_\text{stop}}^t e^{-\delta(t-u)}I(u)\dd{t}
\end{equation}
Finally, a first order estimate of $\widetilde{\Delta S}(I(t),t)$ can be found by exploiting the relation $\Delta S(t) + I(t) + \Delta R(t) = 0$. This, in turn, allows us to numerically estimate both the magnitude of the maximum value reached by the susceptible compartment $S_\text{max}$ as well as the time at which it is reached $t_\text{max}$ by looking for the point in which the condition $\dv{t}(S_0(t)+\widetilde{\Delta S}(I(t),t))=0$ is satisfied, see the Supplementary Material.

\subsection*{Details of the COVID-19 model}
As already mentioned, the COVID-19 model that we used in this work is an extension of the model introduced by Arenas et al.~\cite{Arenas2020} and, therefore, similarly to that model, it has a compartmental dynamics that includes susceptible, exposed, asymptomatic, infected, hospitalized, recovered and deceased, together with some additional compartments to regulate the latency periods. All of these groups are arranged as shown in Fig.~\ref{fig:MMMCA_sketch}. Similarly to the original model, the current version also takes place on a metapopulation network where the mobility is recurrent and modeled through the MIR (Movement-Interaction-Return) framework \cite{gomez2018critical}.

However, differently from the original, the version presented here introduces some features (such as vaccinations and reinfections) that make this model more useful to describe the late stages of the pandemic. In particular, vaccinations have been designed to realistically resemble the mass vaccination campaigns that started in~2021. First of all, unlike all the other transition terms (see Eqs.~(S.28)--(S39) in the Supplementary Material), the rate of vaccination does not depend on the number of unvaccinated individuals, but rather involves a fixed number of vaccines that are distributed among the population according to two rules. The first rule prioritizes patches with a high population density, while the second rule assigns different levels of priority to different age groups according to $\varepsilon^g$. Using these rules, we obtain $\epsilon^g_i(t)$, which represents the absolute number of vaccine doses per age $g$ and location $i$, as follows:
\begin{equation}\label{eq:vac_strategy}
    \epsilon^g_i(t) \propto n_i^g  \cdot  \varepsilon^g\,,
\end{equation}
where the normalization is chosen so that the number of doses per day is fixed. Additionally, we introduced a mechanism to limit and redistribute the doses whenever the number of vaccinated people outnumbers the total number of people in each patch.

From the modeling perspective, being in a vaccinated compartment signifies reduced probabilities of initial infection or, if already infected, of transmitting the disease, being hospitalized, or dying, compared to those in the non-vaccinated compartment. After a certain latency period, individuals lose their vaccine-acquired immunity. However, they transition to a ``post-vaccinated'' status rather than reverting to full susceptibility. In this status, their risk of infection and transmission aligns with that of unvaccinated individuals, but their chances of hospitalization or death remain as low as those who are vaccinated.

Several technical details of our simulations were shared across all simulated scenarios. The specific parameter values used are compiled in Tables~S1 to~S4 in the Supplementary Material, along with the rationale behind their selection. Some parameters were based on our own assumptions or those of previous studies \cite{Arenas2020}, while others were estimated using scientific reports \cite{Informe120, UK_vaccine} and through a calibration process. The initial conditions for our simulations included 10000~individuals in each of the exposed, asymptomatic, and infected compartments. The rest of the population was assumed to be in the susceptible non-vaccinated compartment. The simulations were run for a duration of 600~days, which was chosen to ensure that both the first two infectious peaks were included in the analysis, independently of the vaccination strategy.

\section*{Acknowledgements}

\noindent
 We acknowledge support from Spanish Ministerio de Ciencia e Innovaci\'on (PID2021-128005NB-C21), Generalitat de Catalunya (2021SGR-00633), Universitat Rovira i Virgili (2023PFR-URV-00633) and the European Union’s Horizon Europe Programme under the CREXDATA project (grant agreement no.\ 101092749). AA acknowledges ICREA Academia, the James S.\ McDonnell Foundation (Grant N.\ 220020325), and the Joint Appointment Program at Pacific Northwest National Laboratory (PNNL). PNNL is a multi-program national laboratory operated for the U.S.\ Department of Energy (DOE) by Battelle Memorial Institute under Contract No.\ DE-AC05-76RL01830.

\section*{Contributions}

AA and PC conceived and designed the study. AA and PC designed the methodology and PC performed the analysis. PC, CG, SG and AA drafted the manuscript. All authors edited the manuscript and approved its final version.

\section*{Competing financial interests}

 The authors declare that they have no competing interests. The funding agencies had no role in study design, data collection and analysis, decision to publish, or preparation of the manuscript.

\bibliographystyle{unsrt}

\end{document}


\maketitle

\clearpage

\clearpage

\clearpage

\begin{figure}
    \centering
    \includegraphics[width = 1.0\textwidth]{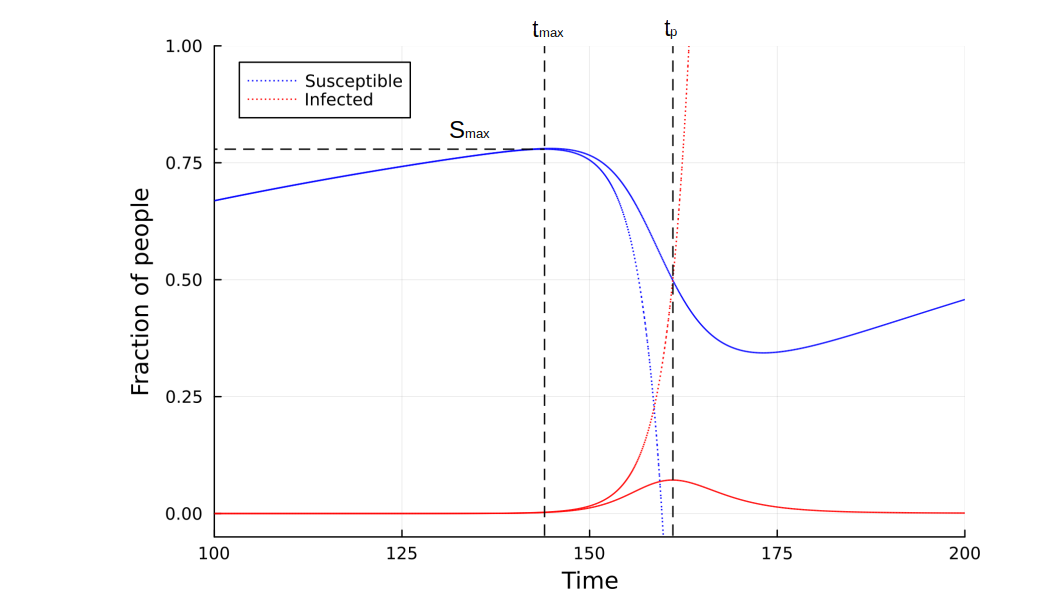}
    \caption{\textbf{No-infected approximation of the SIRS model.} Comparison between the numerical solution of the SIRS model (solid lines) and the analytical formulas (dotted lines). The analytical formulas are the ones displayed in Eqs.~(10)--(12) and obtained in the no-infected approximation. Here, only the susceptible and infected curves are shown. The two vertical lines correspond to $t_\text{max}$ and $t_p$, which are the times at which the maximum of the susceptible curve $S_\text{max}$ is reached and the estimated time of the peak, respectively. Simulation with $\beta=1$, $\mu=0.5$, $\delta=0.01$, $S(0) = 0.1$, $I(0)=10^{-4}$, here displayed only between the times $t=100$ to $t=200$.}
    \label{fig:no-infected-approx}
\end{figure}

\clearpage

\begin{figure}[h]
    \centering
    \includegraphics[width = 0.8\textwidth]{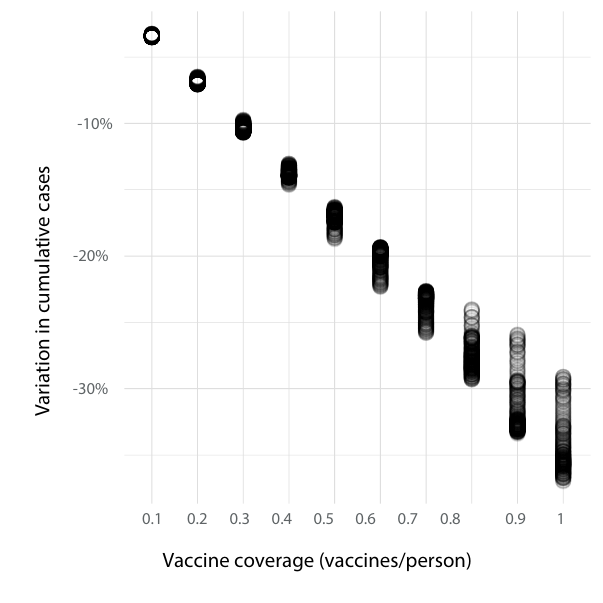}
    \caption{\textbf{Relative reduction in cumulative cases.} Relative reduction in cumulative cases with respect to the scenario with no vaccines as a function of the vaccine coverage, defined as the ratio of vaccines with respect to the total population. Each circle in the figure denotes a specific simulation characterized by a constant vaccine coverage, starting time, and campaign duration. It is evident from the data that vaccine coverage is the primary factor influencing the reduction in total cases. The other two parameters, while less significant, become increasingly noticeable as the vaccine coverage increases. }
    \label{fig:cum_cases}
\end{figure}

\clearpage
\begin{figure*}[tb!]
\begin{center}
\begin{tabular}{cc}
  \includegraphics[width=0.8\textwidth]{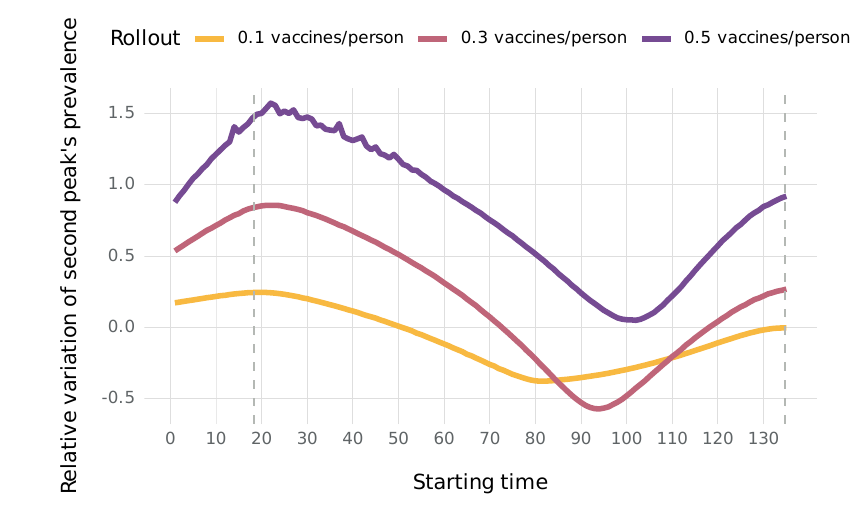} \\
  \includegraphics[width=0.8\textwidth]{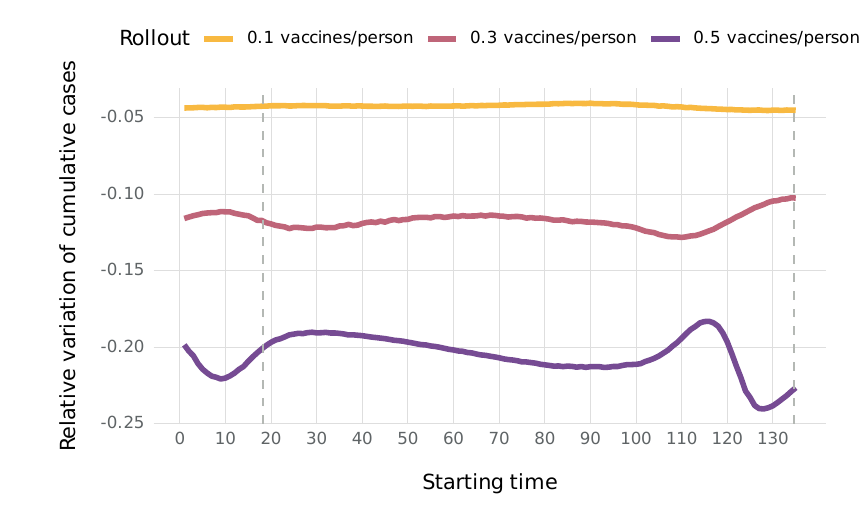}
\end{tabular}
\end{center}
\caption{\textbf{Relative variation of the number of cases at the second peak and cumulative cases due to vaccination strategies with varying durations.} The three vaccination volume levels depicted in this figure correspond to $0.1$, $0.3$ and $0.5$ vaccines per person. The duration of each vaccination campaign is constant and equal to 60~days. The curves have been obtained running the SIRS model with vaccination.
The horizontal axis is the starting day of the vaccination campaign, and the vertical axis is the relative height variation of the second peak of cases (top), and the cumulative number of cases (bottom). The variation is measured relative to the baseline scenario with no vaccines. The dashed vertical lines indicate the location of the first and second peaks in the baseline simulation. Observing the bottom plot, we see that those strategies that deliver more vaccines over time produce better results in terms of containing the number of cumulative cases. However, looking at the top plot, we see that the same longer duration strategies also cause a larger rebound effect. Simulations run with $\beta = 1$, $\mu=0.5$, $\delta=0.01$, $I(0)=10^{-4}$.}
\label{fig_total_vaccines_variation}
\end{figure*}

\clearpage

\begin{figure*}[tb!]
    \centering
    \includegraphics[width = \textwidth]{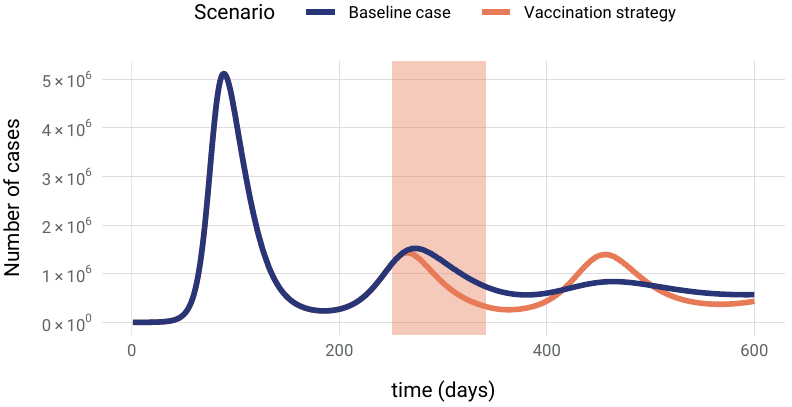}
    \caption{\textbf{Appearance of subsequent peaks due to improper timing of the vaccination strategies.}
    The dark blue line shows the outcome of the baseline scenario where no vaccination campaign is active, while the orange line is the result of a certain vaccination strategy where the timing of the campaign during the second wave exacerbates the third wave. The shaded area denotes the duration of the campaign, which starts at $t=251$ and lasts for 90~days. Here, the age priority vector is set to $[0.1, 0.4, 0.5]$. 
    This example shows that an improperly timed vaccination during the second peak can significantly amplify the following third peak. This observation highlights the generalizability of the conclusions regarding the first and second peaks to any consecutive peaks in the epidemic, emphasizing the importance of well-timed vaccination strategies to mitigate the severity of future waves.
    }
    \label{fig:3rdpeak}
\end{figure*}

\clearpage
\begin{figure*}[tb!]
\begin{center}
  \includegraphics[width=0.9\columnwidth]{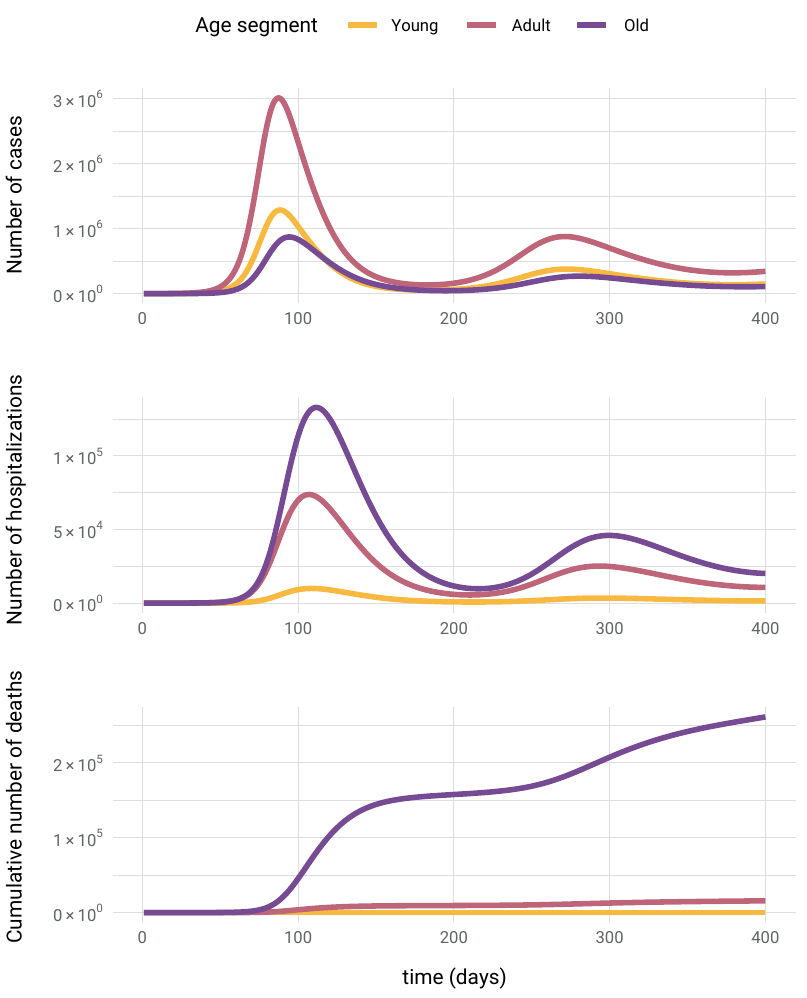}
\end{center}
\caption{\textbf{Number of cases, hospitalizations and cumulative number of deaths in the baseline scenario, for different age strata.} This figure illustrates the breakdown of cases, hospitalizations, and cumulative deaths in the baseline scenario (without an active vaccination campaign) across different age groups. As we can see on the top plot, the adult population contributes most to active cases, followed by the young and elderly. However, hospitalizations (middle plot) predominantly affect the elderly, followed by adults and the young. In terms of fatalities (bottom plot), the elderly group significantly dominates the numbers.}
\label{fig_baseline_by_age}
\end{figure*}

\clearpage
\begin{figure*}[tb!]
    \centering
    \includegraphics[width = \textwidth]{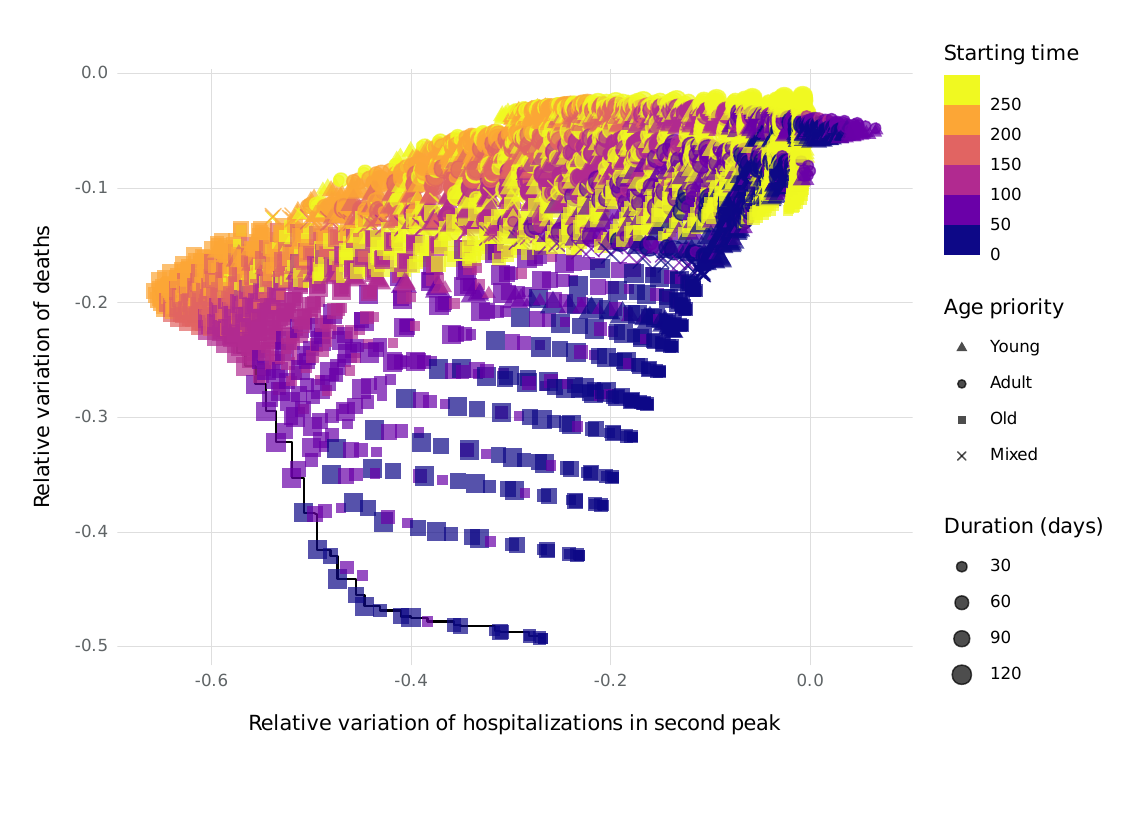}
    \caption{\textbf{Analysis of the vaccination strategies (hospitalizations).} In this scatter plot, each point corresponds to a unique simulation based on a distinct vaccination strategy. The horizontal axis represents the  variation in the number of hospitalizations at the second peak relative to the baseline case, while the vertical axis depicts the relative variation in the number of deaths. Simulation points that fall on positive values of either the horizontal or vertical axis represent counterproductive strategies, that is, they yield a higher number of hospitalizations or deaths compared to the baseline, respectively. The black solid line represents the Pareto front, indicating the optimal strategies that provide the best trade-off between hospitalizations and deaths.}
    \label{fig:pareto}
\end{figure*}

\clearpage

\begin{figure*}[tb!]
    \centering
    \includegraphics[width = \textwidth]{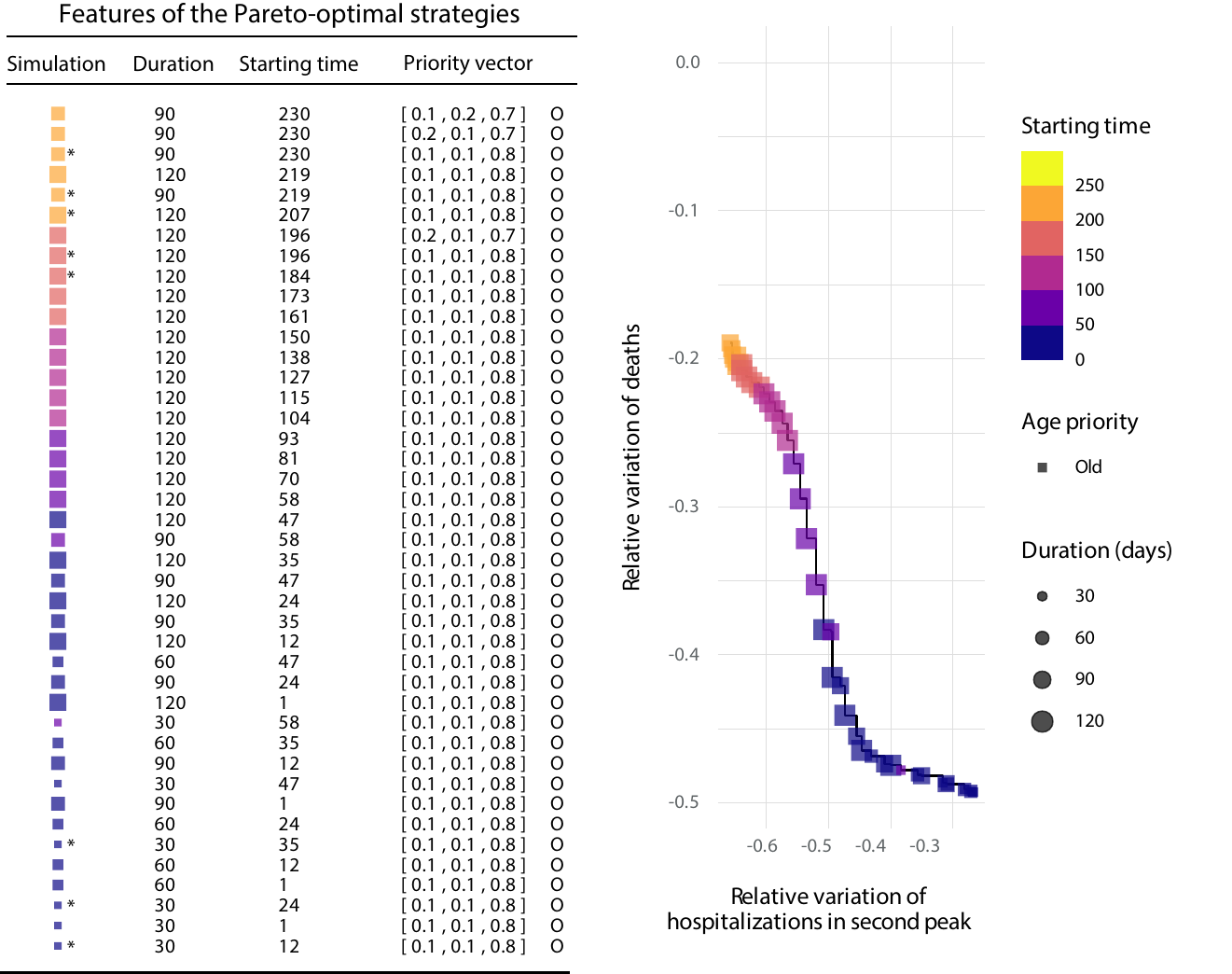}
    \caption{\textbf{Features of the Pareto-optimal strategies (hospitalizations).} This table shows the characteristics of Pareto-optimal strategies alongside a scatter plot that includes only those data points belonging to the Pareto front. The columns of the table provide information on the duration and starting time of each strategy, the priority vector that determines the vaccines distribution by age group (ordered as [Young, Adult, Old]), as well as the age stratum that is most prioritized for each strategy. The asterisk indicates simulations that are optimal for reducing both the number of cases and the number of hospitalizations. As illustrated in the table equivalent to the one presented here, but using cases as the measure (see main text), we can still identify two distinct types of vaccination strategies. However, in this case, the transition between the two kinds is smoother. 
    This can be attributed to the fact that hospitalizations and deaths are more dependent on each other than cases and deaths, and therefore the strategies to optimally reduce one of them are bound to have a certain impact on the other one as well.}
    \label{fig:pareto-front-hosp}
\end{figure*}

\clearpage

\begin{figure*}
    \centering
    \includegraphics[width = 0.7\textwidth]{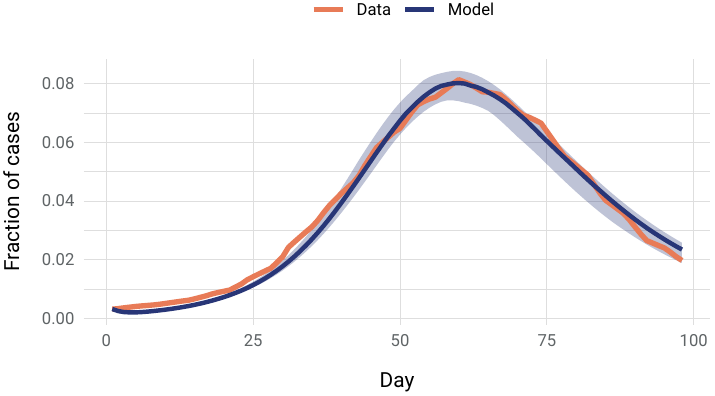}
    \caption{\textbf{Model calibration through the comparison of active cases in Spain.} The time interval considered here goes from the 1st of November 2021 to the 15th of March 2022. The orange line corresponds to real data \cite{worldmeter} while the blue line corresponds to the best model prediction. The shadowed area represents the 95\%~confidence interval.}
    \label{fig:calibration}
\end{figure*}

\clearpage

\section{Rebound timing and its dependence on vaccination}

As we already mentioned in the Methods section of the paper, the solution for the infected individuals after the end of an eradication-type vaccination campaign is:
\begin{equation}
    I(t) = I(t_\text{stop}) \exp\left[ (\beta -\mu) (t-t_\text{stop}) - \frac{\beta R(t_\text{stop})}{\delta}(1- e^{-\delta (t-t_\text{stop})})\right].
\end{equation}
Since it is reasonable to think that the no-infection approximation stops working right around the time the infectious peak is reached, this allows us to use the time $t_p$ where the formula loses meaning as an estimate of the timing of the next epidemic peak. In particular $t_p$ is defined as the time such that $I(t_p)=1-\mu/\beta$, which is an upper-bound for the height of any peak. We know this because Eqs.~(1)--(3) of the SIRS model imply that, when $\dot{I}=0$ (e.g., in a maximum for $I(t)$), the value of the susceptible compartment should be $S=\mu/\beta$, which in turn puts a strong constrain on the sum of the other two compartments.

A closed formula for $t_p$ can be obtained using Lambert $W$ function \cite{corless_lambertw_1996} as follows:
\begin{align}\label{eq:tp_closed_formula}
    t_p = t_\text{stop} +& \dfrac{\beta}{\beta - \mu}\dfrac{R(t_\text{stop})}{\delta} - \dfrac{1}{\beta-\mu}\ln{\dfrac{\beta I(t_\text{stop})}{\beta - \mu}} \notag
    \\
    +&\frac{1}{\delta} W\left( -\dfrac{\beta R(t_\text{stop})}{\beta-\mu} \exp{-\dfrac{\beta R(t_\text{stop})}{\beta - \mu} + \dfrac{\delta}{\beta-\mu}\ln{\dfrac{\beta I(t_\text{stop})}{\beta - \mu}} } \right).
\end{align}
This expression is too complex to be used in regular calculations, but it allows us to understand how the rebound timing $t_p$ depends on the strength of the previous vaccination through its dependencies on $I(t_\text{stop})$ and $S(t_\text{stop})$. We do this by looking at the partial derivatives $\pdv{t_p}{I(t_\text{stop})}$ and $\pdv{t_p}{S(t_\text{stop})}$. We consider $I(t_\text{stop})$ and $S(t_\text{stop})$ as the independent variables, and substitute $R(t_\text{stop})$ using the normalization condition, i.e., $R(t_\text{stop})=1-S(t_\text{stop})-I(t_\text{stop})$. Thus,
\begin{equation}
  \pdv{R(t_\text{stop})}{I(t_\text{stop})}
  =
  \pdv{R(t_\text{stop})}{S(t_\text{stop})}
  =-1\,.
\end{equation}
Furthermore, we will use the following property of the principal branch of the Lambert $W$ function:
\begin{eqnarray}
    \dv{W(x)}{x} = \dfrac{W(x)}{x(W(x)+1)} > 0\,, \quad \forall x>-\frac{1}{e}\,.
\end{eqnarray}

We introduce two auxiliary variables $A$ and $B$, where $A$ corresponds to the argument of the exponential inside the Lambert $W$ function, and $B$ is the argument of the Lambert function itself:
\begin{eqnarray}
    A &=& -\dfrac{\beta R(t_\text{stop})}{\beta - \mu} + \dfrac{\delta}{\beta-\mu}\ln{\dfrac{\beta I(t_\text{stop})}{\beta - \mu}}\,, \\
    B &=& -\dfrac{\beta R(t_\text{stop})}{\beta-\mu} \exp{A} = -\dfrac{\beta R(t_\text{stop})}{\beta-\mu} \exp{-\dfrac{\beta R(t_\text{stop})}{\beta - \mu} + \dfrac{\delta}{\beta-\mu}\ln{\dfrac{\beta I(t_\text{stop})}{\beta - \mu}}}\,,
\end{eqnarray}
thus Eq.~\eqref{eq:tp_closed_formula} can be rewritten as
\begin{eqnarray}
    t_p = t_\text{stop} - \dfrac{A}{\delta} + \dfrac{W(B)}{\delta}\,.
\end{eqnarray}

Before we proceed to calculate the derivatives of the time $t_p$ with respect to $I(t_\text{stop})$ and $S(t_\text{stop})$, it is useful to calculate the derivatives of $A$ and $B$:
\begin{eqnarray} \label{eq:pdvA1}
    \pdv{A}{I(t_\text{stop})} &=& \dfrac{\beta}{\beta-\mu} + \dfrac{\delta}{I(t_\text{stop})(\beta-\mu)}\,,
    \\ \label{eq:pdvA2}
    \pdv{A}{S(t_\text{stop})} &=& \dfrac{\beta}{\beta-\mu}\,,
    \\
    \pdv{B}{I(t_\text{stop})} &=& B \left( \pdv{A}{I(t_\text{stop})} - \dfrac{1}{R(t_\text{stop})} \right),
    \\
    \pdv{B}{S(t_\text{stop})} &=& B \left( \pdv{A}{S(t_\text{stop})} - \dfrac{1}{R(t_\text{stop})} \right).
\end{eqnarray}
Note that Eqs.~\eqref{eq:pdvA1} and~\eqref{eq:pdvA2} are always positive. Armed with these definitions we write the first partial derivative of $t_p$ as follows:
\begin{eqnarray}
    \pdv{t_p}{I(t_\text{stop})} &=& -\dfrac{1}{\delta} \pdv{A}{I(t_\text{stop})} + \dfrac{1}{\delta} \dv{W(B)}{B} \pdv{B}{I(t_\text{stop})}
    \\
    &=& -\dfrac{1}{\delta} \pdv{A}{I(t_\text{stop})} + \dfrac{1}{\delta}\dv{W(B)}{B} B \left( \pdv{A}{I(t_\text{stop})} - \dfrac{1}{R(t_\text{stop})} \right)
    \\
    &=& - \dfrac{1}{\delta} \pdv{A}{I(t_\text{stop})} - \dfrac{1}{\delta} \dv{W(B)}{B} \dfrac{\beta R(t_\text{stop})}{\beta-\mu}\exp{A} \left( \pdv{A}{I(t_\text{stop})} - \dfrac{1}{R(t_\text{stop})} \right).
\end{eqnarray}
We immediately notice that this partial derivative is negative if the term in the parenthesis is positive, which can be written as
\begin{eqnarray}
    \pdv{A}{I(t_\text{stop})} - \dfrac{1}{R(t_\text{stop})} > 0 \quad \Longleftrightarrow \quad R(t_\text{stop}) > \dfrac{(\beta - \mu)I(t_\text{stop})}{\beta I(t_\text{stop}) + \delta}\,.
\end{eqnarray}
Since we are currently working with vaccination, in the eradication regime we can safely assume that $I(t_\text{stop}) \ll 1$ and, therefore, that the above inequality is satisfied and
\begin{equation}
  \pdv{t_p}{I(t_\text{stop})}<0\,.
\end{equation}
On the other hand, the other partial derivative reads
\begin{eqnarray}
    \pdv{t_p}{S(t_\text{stop})} &=& -\dfrac{1}{\delta} \pdv{A}{S(t_\text{stop})} + \dfrac{1}{\delta} \dv{W(B)}{B} \pdv{B}{S(t_\text{stop})}
    \\
    &=& - \dfrac{1}{\delta} \pdv{A}{S(t_\text{stop})} + \dfrac{1}{\delta} \dv{W(B)}{B} B \left( \pdv{A}{S(t_\text{stop})} - \dfrac{1}{R(t_\text{stop})} \right)
    \\
    &=& - \dfrac{1}{\delta} \dfrac{\beta}{\beta-\mu}  - \dfrac{1}{\delta} \dv{W(B)}{B} \dfrac{\beta R(t_\text{stop})}{\beta-\mu}\exp{A} \left( \dfrac{\beta}{\beta-\mu}  - \dfrac{1}{R(t_\text{stop})} \right).
\end{eqnarray}
As before, this expression is negative if the term inside the parenthesis is positive. We can write such condition as
\begin{eqnarray}
    \dfrac{\beta}{\beta-\mu}  - \dfrac{1}{R(t_\text{stop})}  > 0 \quad \Longleftrightarrow \quad S(t_\text{stop}) < \dfrac{\mu}{\beta}\,,
\end{eqnarray}
which is always true in the eradication regime, therefore proving that, in this scenario,
\begin{equation}
  \pdv{t_p}{S(t_\text{stop})}<0\,.
\end{equation}

To sum up, what we have found is that, the smaller $I(t_\text{stop})$ and $S(t_\text{stop})$, the larger $t_p$ will be. Furthermore, since in the eradication regime both of these quantities are related to the rate of vaccination $\alpha$, what we have demonstrated here is that, the stronger the vaccination campaign (i.e., the larger $\alpha$), the more delayed the rebound will be.

\section{Rebound height and its dependence on vaccination}
Finding a good way of estimating the height of an epidemic peak is challenging, and every estimate ends up relying on a number of uncontrolled assumptions. However, in our case we are not so much interested in its numerical estimation, but in finding a formula that displays a similar qualitative behavior.

Our estimate of the epidemic peak relies on the connection between it and the slope of the susceptible curve at time $t_p$. The two are connected by the Eq.~(1) in the following way:
\begin{eqnarray}
    \dot{S}(t_p) = - \beta S(t_p) I(t_p) + \delta R(t_p)\,.
\end{eqnarray}
By using the facts that $S(t_p)=\mu/\beta$ and $R(t)=1-S(t)-I(t)$, rearranging the terms we get that:
\begin{eqnarray}\label{eq:Ip-estimate}
    I(t_p) = I^* - \dfrac{\dot{S}(t_p)}{\mu + \delta}\,,
\end{eqnarray}
where $I^* = \frac{\delta}{\mu+\delta}\left(1 - \frac{\mu}{\beta}\right)$ is the equilibrium value of the epidemic compartment in the SIRS model without vaccination. Notice that Eq.~\eqref{eq:Ip-estimate} holds true for every maximum and minimum of the function $I(t)$, but we are here only interested in the first maximum.

The problem now becomes to find the slope of the function $S(t)$ at time $t_p$. We achieve this with a quadratic approximation $\hat{S}(t) = at^2 + bt + c$ that satisfies the following three conditions:
\begin{eqnarray} \label{eq:condition_parabola}
    \hat{S}(t_\text{max}) = S_\text{max}, \qquad \hat{S}(t_p) = \dfrac{\mu}{\beta}, \qquad \hat{S}'(t_\text{max}) = 0,
\end{eqnarray}
where $S_\text{max}$ is the maximum value reached by the susceptible compartment before it starts to decrease, while $t_\text{max}$ is the time at which that happens. In our approximation we make this point coincide with the vertex of the parabola. The three conditions in Eq.~\eqref{eq:condition_parabola} give us a system of three equations with three unknowns, i.e., the parameters of the parabola. It is therefore always possible to find an explicit solution:
\begin{eqnarray}
    a = -\dfrac{S_\text{max} - \mu/\beta}{(t_p-t_\text{max})^2}, \qquad     b = -2 t_\text{max} a, \qquad c =  S_\text{max} + a t_\text{max}^2\,.
\end{eqnarray}
Next, we can simply compute the derivative of the parabola at time $t_p$, $\hat{S}(t_p)=2a t_p+b$, which gives us an estimate for the slope of the susceptible curve at that point
\begin{eqnarray}
    \dot{S}(t_p) = - 2 \dfrac{S_\text{max} - \mu/\beta}{t_p-t_\text{max}}\,.
\end{eqnarray}
Finally, thanks to Eq.~\eqref{eq:Ip-estimate}, we get an estimate for the height of the peak as a function of the peak in susceptibles preceding the rebound
\begin{eqnarray}
    I(t_p) = I^* + \dfrac{2(S_\text{max} - \mu/\beta)}{(\mu+\delta)(t_p-t_\text{max})}\,.
\end{eqnarray}

The conclusion we can draw from this estimation is that, the larger the build-up in susceptibles before an epidemic wave (i.e., $S_\text{max}$), the taller the subsequent peak will be.

\section{Equations of the age-stratified COVID-19 model}

The variables of our system of equations are \{$\rho^{m,g}_{i,v}(t)$\}, which indicate the density of individuals in the compartment~$m$, age stratus~$g$, location~$i$ and vaccination status~$v$ at time~$t$ (measured in days). In the following, we write the equations for the temporal evolution of these quantities. Note that, for the majority of these compartments, the indexes~$g$, $i$ and~$v$ can be left implicit, while in the case of the compartment~$S$, the index~$v$ must be specified because the form of the equation changes according to it. Apart from adding the vaccination state, only the equations for compartments~S and~R have been changed with respect to the work by Arenas et al.~\cite{Arenas2020}:
\begin{eqnarray}
  \rho^{S,g}_{i,0}(t+1) &=&  (1-\Pi^{g}_{i, 0}(t)) \rho^{S,g}_{i,0}(t) - \epsilon^g_i / n^g_i\,,
  \label{eq:modelS0}
  \\
  \rho^{S,g}_{i,1}(t+1) &=&  (1-\Pi^{g}_{i, 1}(t)) \rho^{S,g}_{i,1}(t) - \Lambda \rho^{S,g}_{i,1}(t) + \epsilon^g_i(t) / n^g_i\,,
  \label{eq:modelS1}
  \\
  \rho^{S,g}_{i,2}(t+1) &=&  (1-\Pi^{g}_{i, 2}(t)) \rho^{S,g}_{i,2}(t) + \Lambda \rho^{S,g}_{i,1}(t) + \Gamma (\rho^{R,g}_{i,0}(t) + \rho^{R,g}_{i,1}(t) + \rho^{R,g}_{i,2}(t) )\,,
  \label{eq:modelS2}
  \\
  \rho^{E,g}_{i, v}(t+1) &=& (1 - \eta^g)\rho^{E,g}_{i, v}(t) + \Pi^{g}_{i, v}(t) \rho^{S,g}_{i, v}(t)\,,
  \label{eq:modelE}
  \\
  \rho^{A,g}_{i, v}(t+1) &=& (1-\alpha^g)\,\rho^{A,g}_{i, v}(t) + \eta^g\,\rho^{E,g}_{i, v}(t)\,,
  \label{eq:modelA}
  \\
  \rho^{I,g}_{i, v}(t+1) &=& (1-\mu^g)\,\rho^{I,g}_{i, v}(t) + \alpha^g\,\rho^{A,g}_{i, v}(t)\,,
  \label{eq:modelI}
  \\
  \rho^{P_D,g}_{i, v}(t+1) &=& (1-\zeta^g)\,\rho^{P_D,g}_{i, v}(t) + \mu^g\,\theta_v^g\,\rho^{I,g}_{i, v}(t)\,,
  \label{eq:modelLD}
  \\
  \rho^{P_H,g}_{i, v}(t+1) &=& (1-\lambda^g)\,\rho^{P_H,g}_{i, v}(t) + \mu^g\,(1-\theta_v^g)\,\gamma_v^g\,\rho^{I,g}_{i, v}(t)\,,
  \label{eq:modelLH}
  \\
  \rho^{H_D,g}_{i, v}(t+1) &=& (1-\psi^g)\,\rho^{H_D,g}_{i, v}(t) + \lambda^g\,\omega^g_v\,\rho^{P_H,g}_{i, v}(t)\,,
  \label{eq:modelHD}
  \\
  \rho^{H_R,g}_{i, v}(t+1) &=& (1-\chi^g)\,\rho^{H_R,g}_{i, v}(t) + \lambda^g\,(1-\omega^g_v)\,\rho^{P_H,g}_{i, v}(t)\,,
  \label{eq:modelHR}
  \\
  \rho^{D,g}_{i, v}(t+1) &=& \rho^{D,g}_{i, v}(t) + \zeta^g\,\rho^{P_D,g}_{i, v}(t) + \psi^g\,\rho^{H_D,g}_{i, v}(t)\,,
  \label{eq:modelD}
  \\
  \rho^{R,g}_{i, v}(t+1) &=& \rho^{R,g}_{i, v}(t) + \mu^g\,(1-\theta_v^g)\,(1-\gamma_v^g)\,\rho^{I,g}_{i, v}(t) + \chi^g\,\rho^{H_R,g}_{i, v}(t) - \Gamma \rho^{R,g}_{i,v}(t)\,.
  \label{eq:modelR}
\end{eqnarray}
The following normalization relations hold:
\begin{equation}
    \sum_m\sum_{v\in\{0,1,2\}} \rho^{m,g}_{i, v}(t)=1, \quad \forall g,i,t,
\end{equation}
which can be easily checked by looking at the system of Eqs.~\eqref{eq:modelS0} to~\eqref{eq:modelR}. 

Let us describe the compartmental dynamics of our model. The three vaccination statuses, denoted by ${0,1,2}$, represent the natural progression of vaccine-associated defense changes within an individual. Respectively, they correspond to: a completely defenseless person; a person with high defenses against both infection and hospitalization; and finally, a person with high defenses against hospitalization but who remains totally susceptible to infection. Initially, a fraction~$\epsilon^g_i / n^g_i$ of the susceptible population is vaccinated. Although vaccinated individuals will still go through the same epidemiological compartments, their transition rates will differ based on their vaccination status. We assume that vaccine-induced immunity diminishes with rate~$\Lambda$.
Susceptible individuals become infected when they come into contact with asymptomatic or infected individuals, with the probability of transmission denoted by $\Pi^g_{i,v}$ (further explained in the next section). If transmission occurs, the previously susceptible individual moves to the exposed compartment. An exposed individual transitions to the asymptomatic compartment with rate~$\eta^g$ and subsequently to the infected compartment with rate~$\alpha^g$.
From the infected compartment, several outcomes are possible, that are reached at a rate~$\mu^g_v$. One possibility is recovering after the infection without hospitalization, with probability $(1-\theta_v^g)\,(1-\gamma_v^g)$. Another possibility is having a severe course of infection and requiring hospitalization, with probability $(1-\theta_1^g)\gamma_1^g$ and a delay governed by rate $\lambda^g$. At this point, individuals can either receive a fatal prognosis with probability~$\omega^g_v$ leading to death at a rate~$\psi^g$, or a good prognosis with probability $1-\omega^g_v$, leading to recovery at a rate~$\chi^g$. The last possibility is dying without being hospitalized with probability~$\theta^g_v$ after a latency period governed by rate~$\zeta^g$.
Finally, individuals in the recovered compartment who are not vaccinated might transition again to the susceptible compartment with probability~$\Gamma$. All the numerical values of these parameters can be found in Tables~\ref{tab:parameters_epi}, \ref{tab:parameters_clin} and~\ref{tab:parameters_vac}.

\section{Social contacts and infection probability}

It is important to specify how the infection probability $\Pi^{g}_{i, v}(t)$ is calculated. This probability represents the probability of an individual associated to patch~$i$, age stratum~$g$, and vaccination status~$v$ to be infected at time~$t$. Following Arenas et al.~\cite{Arenas2020}, the infection probability is calculated as
\begin{equation}
    \Pi^{g}_{i, v}(t) = (1 - p^g) P^g_{i, v}(t) + p^g \sum_j R_{ij}^g P_{j, v}^g(t)\,,
\end{equation}
where the first term indicates the probability of susceptible individuals to get infected in their ``home'' patch while the second term indicates the probability of getting infected elsewhere. In particular $p^g$ is the probability to travel from your patch to another and $R_{ji}^g$ is the probability that an individual of age $g$ will go from $i$ to $j$, given that it will move (no correlation is assumed between the vaccination status $v$ and the mobility of an individual). Furthermore $P_{i, v}^g(t)$ denotes the probability that an agent of age~$g$ and status~$v$ gets infected inside patch~$i$. This is, in turn, expressed as
\begin{eqnarray}
    P_{i, v}^g(t) &=& 1 - \prod_{h,j,w} \prod_{m\in\{A,I\}} \bigg[ 1 - \beta^m (1 - r_v)(1-b_w)\bigg]^{ T_{j\to i}^{m, h, w} },
\end{eqnarray}
where $r_v$ is the vaccine efficacy in preventing infections while $b_w$ is the vaccine efficacy in preventing transmission once already infected. The exponent $T_{j\to i}^{m, h, w}$ indicates the effective number of contacts made by an agent of age~$h$, compartment~$m$ (either infected of asymptomatic) and vaccination status~$w$ that traveled from patch~$j$ to patch~$i$. This quantity is calculated as follows:
\begin{equation}
    T_{j\to i}^{m, h, v} = z^g \expval{k^g}f(n^g_i/s_i) C^{gh} \dfrac{n_{j\to i}^{m, h, v}}{\tilde{n}_i^h}\,.
\end{equation}
Here, the term $z^g \expval{k^g}f(n^g_i/s_i)$ represents the total number of contacts that people of age~$g$ make inside patch~$i$. Those contacts increase monotonically with the population density in that patch, and the function that we use to model this dependency is the following~\cite{Arenas2020}:
\begin{equation}
    f(x) = 1 + (1-e^{-\xi x})\,.
\end{equation}
Since we want the overall number of contacts to depend on the average number of connections of each age group, we introduce~$\expval{k^g}$ multiplied by a normalization factor~$z^g$ such that the average degree of population belonging to age group~$g$ is exactly~$\expval{k^g}$. From \cite{Arenas2020} we know that:
\begin{equation}
    z^g = \dfrac{n^g}{\sum_{i=1}^{N_P}f(\frac{\tilde{n}_i}{s_i})\tilde{n}^g_i}\,,
\end{equation}
where $s_i$ is the surface of the patch~$i$, and the symbols~$\tilde{n}_i$ and~$\tilde{n}_i^g$ refer to the number of people present in patch~$i$ during the commuting phase, given by:
\begin{equation}
    \tilde{n}_i = \sum_{g=1}^{N_G}\tilde{n}_i^g
\end{equation}
and
\begin{equation}
    \tilde{n}_i^g = \sum_{j=1}^{N_P} M_{ji}^g n^g_j\,,
\end{equation}
For convenience, we also define the mobility matrix as
\begin{equation}
    M_{ji}^g = (1-p^g)\delta_{ji} + p^gR_{ji}^g\,,
\end{equation}
being $\delta_{ji}$ the Kronecker delta function. Then, the total number of contacts must be multiplied by~$C^{gh}$, which specifies the fraction of all the contacts that individuals in age group~$g$ have with individuals in age group~$h$. Finally, the last term in the exponent indicates how many of these contacts were with infected or asymptomatic people coming from node $j$:
\begin{equation}
    n_{j\to i}^{m, h, v}(t) = n_j^h \rho^{m,h}_{j,v}(t) M_{ji}^h\,.
\end{equation}

All the numerical values of the parameters in this section can be found in Table \ref{tab:parameters_pop}.

\clearpage

\begin{table*}[tb!]
    \centering
    \begin{tabular}{lccc}
    \hline
    Symbol & Description & Estimates in Spain & Assignment \\
    \hline
        $\beta^I$ & Infectivity of symptomatic & 0.056 & Calibrated \\
        $\beta^A$ & Infectivity of asymptomatic & $\beta^I/2$ & Assumed \\
        $\eta^g$ & Exposed rate & 0.127 & Calibrated\\
        $\alpha^g$ & Asymptomatic rate & 0.306 & Calibrated \\
        $\mu^g$ & Infectious rate & 0.589 & Calibrated \\
    \hline
    \end{tabular}
    \caption{\textbf{Epidemic parameters.} Parameters of the epidemic, determined through calibration with real data on the number of active cases of COVID-19 in Spain between the 1st of December 2021 and the 15th of March 2022 \cite{worldmeter}, see Fig.~\ref{fig:calibration}. The calibration was carried out using the Turing package of the Julia language, which relies on a Markov Chain Monte Carlo approach with a No-U-Turn sample \cite{hoffman2014no}.}
    \label{tab:parameters_epi}
\end{table*}

\begin{table*}[tb!]
    \centering
    \begin{tabular}{lccc}
    \hline
    Symbol & Description & Estimates in Spain & Assignment \\
    \hline
        $\theta^g_v$ & Direct death probability & $0.0 $ & \cite{Informe120} \\
        \{$\gamma^g_v$\} & ICU probability & $(0.003, 0.01,0.08)^g\otimes (0,0.15)_v$ & \cite{Informe120, UK_vaccine} \\
        \{$\omega^g_v$\} & Death probability in ICU & $(0, 0.04, 0.3)^g\otimes (0,0.1)_v$ & \cite{Informe120, UK_vaccine} \\
        $\lambda^g$ & Prehospitalized in ICU rate & 4.084 days$^{-1}$ & \cite{Arenas2020} \\
        $\zeta^g$ & Predeceased rate & 7.084 days$^{-1}$ & \cite{Arenas2020} \\
        $\psi^g$ & Death rate in ICU & 7 days$^{-1}$ & \cite{Arenas2020} \\
        $\chi^g$ & ICU discharge rate & 21 days$^{-1}$ & \cite{Arenas2020} \\
    \hline
    \end{tabular}
    \caption{\textbf{Clinical parameters.} Clinical parameters, taken from Arenas et al. \cite{Arenas2020}.}
    \label{tab:parameters_clin}
\end{table*}

\begin{table*}[tb!]
    \centering
    \begin{tabular}{lccc}
    \hline
    Symbol & Description & Estimates in Spain & Assignment \\
    \hline
        $\Gamma$ & Reinfection rate & 100 days$^{-1}$ & Assumed \\
        $\Lambda$ & Waning immunity rate & 50 days$^{-1}$  & Assumed \\
        \{$r_v$\} & Risk Reduction of infection probability & $(0.0, 0.6)$ & \cite{UK_vaccine} \\
        \{$b_v$\} & Risk Reduction of transmission probability & $(0.0, 0.4)$ & \cite{UK_vaccine} \\
    \hline
    \end{tabular}
    \caption{\textbf{Vaccination and immunity parameters.} Selected vaccination and immunity parameters.}
    \label{tab:parameters_vac}
\end{table*}

\begin{table*}[tb!]
    \centering
    \begin{tabular}{lcc}
    \hline
    Symbol & Description & Estimates for $g\in\{Y,M,O\}$ in Spain \\
    \hline
        \{$N^g$\} & Population by age stratum & (12M, 26,4M, 8,9M) \cite{INE}\\
        $n^g_i$ & Region population & \cite{population_surface} \\
        $s_i$ & Region surface & \cite{population_surface} \\
        $R_{ij}^g$ & Mobility matrix (non-diagonal) & \cite{mitba} \\
        $\expval{k^g}$ & Average total number of contacts & \cite{Arenas2020} \\
        $C^{gh}$ & Contacts-by-age matrix & \cite{Arenas2020} \\
        $\xi$ & Density factor & $0.01\ \text{km}^2$ \cite{Arenas2020} \\
        \{$p^g$\} & Mobility factor & $(0.3, 1.0, 0.05)$ \cite{Arenas2020} \\
    \hline
    \end{tabular}
    \caption{\textbf{Social parameters.} Parameters of the model related to geographic and population data, including mobility, and their values for Spain.}
    \label{tab:parameters_pop}
\end{table*}

\clearpage

\bibliographystyle{unsrt}